\documentclass[aps,prb,amsmath,amssymb,reprint,superscriptaddress]{revtex4-2}

\usepackage{mathrsfs,amsfonts} 

\usepackage{bbm}
\usepackage[colorlinks,citecolor=blue,linkcolor=red,urlcolor=blue]{hyperref}
\usepackage{color}
\usepackage{graphicx}
\usepackage{subfigure}
\usepackage{verbatim}
\usepackage{fourier}
\usepackage{soul} 

\begin{document}

\title{Near-field radiative heat transfer between shifted graphene gratings}

\author{Minggang Luo}
\email[]{minggang.luo@umontpellier.fr}
\affiliation{Laboratoire Charles Coulomb (L2C), UMR 5221 CNRS-Universit\'e de Montpellier, F-34095 Montpellier, France}

\author{Youssef Jeyar}
\affiliation{Laboratoire Charles Coulomb (L2C), UMR 5221 CNRS-Universit\'e de Montpellier, F-34095 Montpellier, France}

\author{Brahim Guizal}
\affiliation{Laboratoire Charles Coulomb (L2C), UMR 5221 CNRS-Universit\'e de Montpellier, F-34095 Montpellier, France}


\author{Mauro Antezza}
\email[]{mauro.antezza@umontpellier.fr}
\affiliation{Laboratoire Charles Coulomb (L2C), UMR 5221 CNRS-Universit\'e de Montpellier, F-34095 Montpellier, France}
\affiliation{Institut Universitaire de France, 1 rue Descartes, Paris Cedex 05, F-75231, France}

\date{\today}

\begin{abstract}

We examine the near-field radiative heat transfer between finite-thickness planar fused silica slabs covered with graphene gratings, through the utilization of the Fourier modal method augmented with local basis functions (FMM-LBF), with focus on the lateral shift effect. To do so, we propose and validate a minor modification of the FMM-LBF theory to account for the lateral shift. This approach goes far beyond the effective medium approximation because this latter cannot account for the lateral shift. 
We show that the heat flux can exhibit significant oscillations with the lateral shift and, at short separation, it can experience up to a 60\%-70\% reduction compared to the aligned case.
Such a lateral shift effect is found to be sensitive to the geometric factor $d/D$ (separation distance to grating period ratio). When $d/D>1$ (realized through large separation or small grating period), the two graphene gratings see each other as an effective whole rather than in detail, and thus the lateral shift effect on heat transfer becomes less important. Therefore, we can clearly distinguish two asymptotic regimes for radiative heat transfer: the LSE (Lateral Shift Effect) regime, where a significant lateral shift effect is observed, and the non-LSE regime, where this effect is negligible.
Furthermore, regardless of the lateral shift, the radiative heat flux shows a non-monotonic dependence on the graphene chemical potential. That is, we can get an optimal radiative heat flux (peaking at about 0.3eV chemical potential) by \textit{in situ} modulating the chemical potential.
{This work has the potential to unveil new avenues for harnessing the lateral shift effect on radiative heat transfer in graphene-based nanodevices. }

\end{abstract}

\maketitle 

\section{introduction}
Recently, there has been a growing interest in near-field radiative heat transfer (NFRHT), motivated by both fundamental explorations and practical applications.
%
When the separation distance is comparable to or smaller than the thermal wavelength $\lambda_T=\hbar c/k_BT$, the radiative heat flux can exceed the Planckian blackbody limit by several orders of magnitude \cite{Rytov1989,Polder1971}, due to the near-field effects (\textit{e.g.}, photon tunneling) \cite{Chapuis2008plate,Narayanaswamy2008,Carminati1999,Loomis1994}.
 The NFRHT has been extensively investigated theoretically for many different geometric configurations \cite{Shchegrov2000,Volokitin2001,Chapuis2008,Manjavacas2012,Nikbakht2018,Messina2018,DongPrb2018,Zhang2019T,Luo2020prb_ensemble,Luo2023IJHMT_diffusion,Biehs2011gratings,Kan2019prb,Zheng2022Materials_EMA}, some of which have been confirmed by pioneering experimental works \cite{Shen2009,Rousseau2009,Ottens2011,Song2015,Lim2015,Watjen2016,Ghashami2018,Yungui2018NC,DeSutter2019,Yungui2023PRAppl}.
Particularly, when grating structures are involved, they can lead to important enhancement of the NFRHT due to the excitation of high-order diffraction channels \cite{Yang2017prl,Messina2017PRB,Hongliang_JHMT}. Besides, owing to graphene's special optical properties, planar graphene-sheet involved structures exhibit many novel behaviors in the radiative heat transfer modulation \cite{Svetovoy2012prb,Ognjen2012prb,Zheng2017,Volokitin2017Dey,Zhao2017prb,Shi2021am,LiuES2022,Lu2022small,JSWM,GrafBilMA}. It is interesting to know whether the combined richness of the grating geometry and the special dielectric features of graphene could lead to some other novel behavior in NFRHT manipulation that cannot be achieved with conventional materials.

Radiative energy transfer between graphene gratings based structures \cite{Liu2015APL,Yang2020prAppl,Luo2023apl_NFRHT_grating} has recently been investigated, where patterning the graphene sheets  has been found to open more energy transfer channels. All of these works consider two perfectly aligned and identically patterned graphene structures. The effect of twisted grating has also been investigated \cite{He2020ol,He2020ijhmt,Luo2020apl}. Still, the effect of a lateral shift between the two parallel gratings remains unexplored. Introducing a lateral shift will allow a modulation of the near-field radiative heat transfer keeping constant the separation distance. It is worth stressing that the effective medium theory (EMT), that is often used to study the NFRHT between substrate-supported graphene strips \cite{Zhou2022langmuir}, treats the graphene grating as an effective whole and thus cannot account for the lateral shift effect. The study of a shifted configuration needs {a more accurate} numerical method to treat the electromagnetic wave scattering by the gratings.

%
\begin{figure} [h]
\centerline {\includegraphics[width=0.35\textwidth]{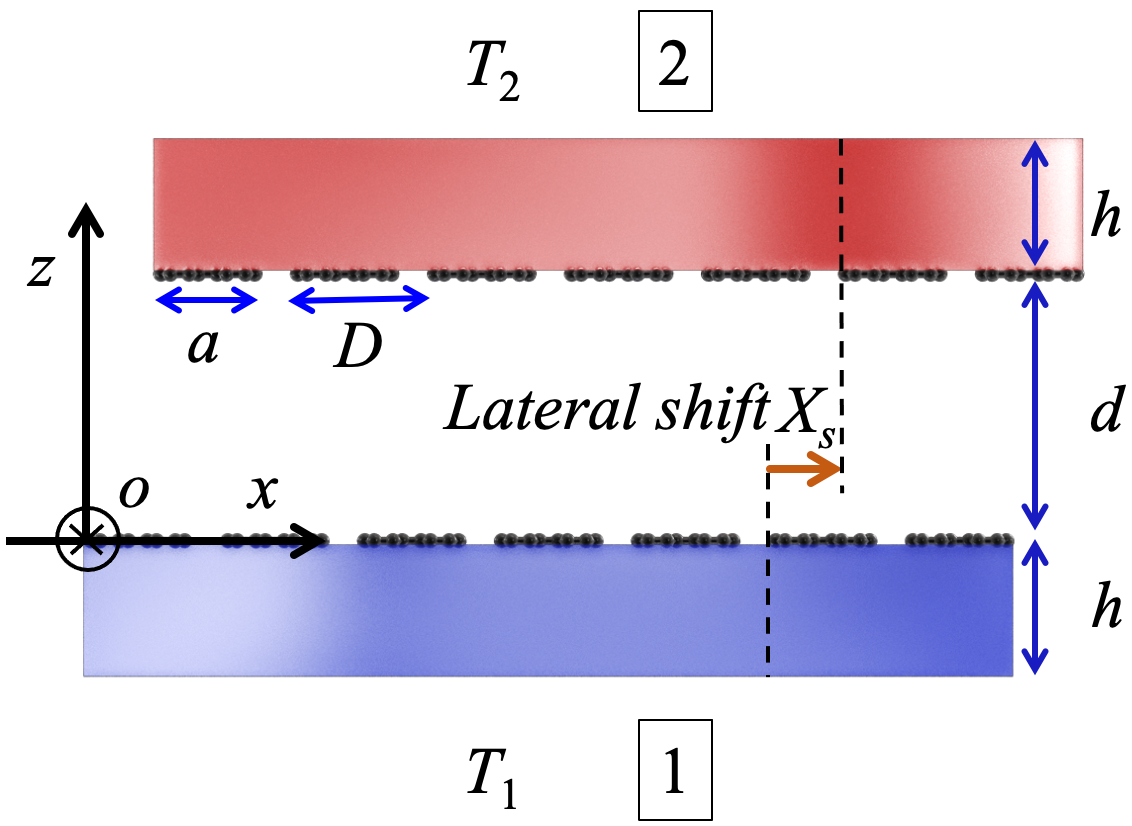}}
     \caption{Structure diagram of the supported graphene gratings considering relative lateral shift $X_s$.}
   \label{fig:two_gratings_schematic}
\end{figure}

In this work we investigate the effect of a lateral shift on NFRHT. We will consider planar fused silica slabs coated with graphene gratings (see Fig.~\ref{fig:two_gratings_schematic}).  Concerning the numerical investigations, we will use {a more accurate} method: the Fourier modal method equipped with local basis functions (FMM-LBF), which is particularly efficient for one-dimensional strip gratings \cite{Guizal, Taiwan_LBF, PRE23,Luo2023Casimir_gg}. 
{Such method does not use any approximations other than the classical ones (harmonic plane wave expansion, linear homogeneous and isotropic materials, ...) in the solution of Maxwell's equations. The `restriction', if it is one, is the truncation necessary for the numerical implementation. There is no such restriction on the wavelength to period ratio as the one we have with the EMT.}
%
In order to take into account this lateral shift, {one can either include it by directly applying the translation to the scattering operator (of non-translated objects), or by including it in the construction of the scattering operators themselves. In our computations, we checked the equivalence of these two approaches.}

This work is organized as follows: In Sec.~\ref{models}, the physical model of the planar fused silica slabs covered with graphene gratings with a relative lateral shift and the theoretical models of the scattering approach for radiative heat transfer are presented. In Sec.~\ref{results_discussion}, the effects of influencing factors (\textit{e.g.}, grating period, chemical potential and filling fraction) on the lateral shift mediated NFRHT are analyzed, and asymptotic regimes of the lateral shift effect are also proposed.

\section{Theoretical models}
\label{models}
%
The physical system is shown in Fig.~\ref{fig:two_gratings_schematic}. We investigate the NFRHT between bodies 1 and 2. The relative lateral displacement between the two graphene strips along the $x$-axis is $X_s$. {When $X_s=0$, the two graphene gratings are perfectly aligned and the near-field radiative heat transfer, in this case, has been recently investigated (please refer to our recent work for more details \cite{Luo2023apl_NFRHT_grating}). }  

In the following, we will denote by $D$ the graphene grating period, by $a$ the width of one single graphene strip, by $h$ the common thickness of the planar slabs and by $d$ the separation distance between the two gratings. Bodies 1 and 2 are considered at fixed temperatures $T_1$ and $T_2$, respectively, while the environment is at the same temperature as body 1.  The electromagnetic properties of graphene are taken into account through its conductivity $\sigma_g$ which depends on the temperature $T$ and on the chemical potential $\mu$. It is the sum of an intraband and an interband contributions $\left(\sigma_g = \sigma_{\textnormal{intra}} +\sigma_{\textnormal{inter}}\right)$  given by \cite{Zhao2017prb,Falkovsky2007,Falkovsky2008,Awan_2016}:
\begin{equation}
\sigma_{\textnormal{intra}} = \frac{i}{\omega + i/\tau} \frac{2 e^2 k_{\rm B} T}{\pi\hbar^2}\ln\left[ 2 \cosh\left(\dfrac{\mu}{2k_{\rm B}T}\right) \right],
\end{equation}
and
\begin{equation}
\sigma_{\textnormal{inter}}= \frac{e^2}{4 \hbar} \left [ G\left(\frac{\hbar \omega}{2}\right) + i \dfrac{ 4\hbar \omega}{\pi}\int_0^{+\infty} \dfrac{G(\xi)-G\left(\hbar \omega /2\right)}{(\hbar \omega)^2 - 4\xi^2} {\rm d}\xi \right],
\end{equation}
where  $e$  is the electron charge, $\tau$ the relaxation time (we use $\tau = 10^{-13}${s}) and $G(\xi) = \sinh (\xi/k_B T) / [\cosh (\mu/k_{\rm B} T) + \cosh(\xi/k_{\rm B} T)]$.

The net power flux $\varphi$ received by body 1 (energy per unit surface and time) can be defined as \cite{Messina2014,Messina2011PRA}
%
\begin{equation}
\varphi= \sum_p \int \frac{{\rm d}^2 {\rm \textbf{k}}}{(2\pi)^2} \int_0^{+\infty} (\Theta(\omega,T_2)-\Theta(\omega,T_1)) \frac{{\rm d} \omega}{2\pi}<p,{\rm \textbf{k}}|\mathcal{O}|p,{\rm \textbf{k}}> ,
\label{HF}
\end{equation}
%
where  $p$ is the polarization index, $p=$ 1, 2 corresponding to TE (transverse electric) and TM (transverse magnetic) polarization modes respectively, $\Theta(\omega,T)=\hbar \omega/( e^{{\hbar \omega}/{k_{\rm B} T}}-1 )$ is the mean energy of the Planck oscillator, $\hbar$ is the reduced Planck constant, $\omega$ is the angular frequency, $k_{\rm B}$ is the Boltzmann constant, ${\rm \textbf{k}}=(k_x, k_y)$, $k_x$ and $k_y$ being the wave vectors in the standard $(x, y, z)$ cartesian coordinates system. The transmission operator $ \mathcal{O}$ in the (TE, TM) basis is given by \cite{Messina2011PRA}
\begin{widetext}
\begin{equation}
\mathcal{O}=   U^{(2,1)} \left[  f_{-1}(\mathcal{R}^{(2)-}) - \mathcal{T}^{(2)-} \mathcal{P}_{-1}^{({\rm pw})}  \mathcal{T}^{(2)-\dagger} \right] U^{(2,1)\dag } \left[  f_{1}(\mathcal{R}^{(1)+}) - \mathcal{T}^{(1)-\dagger} \mathcal{P}_{1}^{({\rm pw})}  \mathcal{T}^{(1)-} \right]  ,
\label{O_operator}
\end{equation}
\end{widetext}
where $U^{(2,1)}=\left(1-\mathcal{R}^{(2)-}\mathcal{R}^{(1)+}\right)^{-1}$, $\mathcal{R}^{(1)+}$ and $\mathcal{R}^{(2)-}$ ($\mathcal{T}^{(1)-}$ and $\mathcal{T}^{(2)-}$ ) are the reflection operators (transmission operators) of grating 1 and grating 2 in the (TE, TM) basis, $\dagger$ stands for the conjugation operation and the $\pm$ superscripts in the reflection and transmission coefficients correspond to the propagation direction with respect to the $z$ axis. $\left \langle p,{\textbf{k}}|\mathcal{P}_{\zeta}^{\rm (pw/ew)}|p',{\textbf{k}}' \right\rangle 
 =k_{z}^{\zeta}	\left \langle p,{\textbf{k}}|\mathcal{\prod}^{\rm (pw/ew)}|p',{\textbf{k}}' \right\rangle$, $k_{z}=\sqrt{k_0^2-\textbf{k}^2}$, $k_0=\omega/c$, $\mathcal{\prod}^{\rm (pw)}$ ($\mathcal{\prod}^{\rm (ew)}$) is the projector on the propagative (evanescent) sector, and the auxiliary function $f_{\alpha}(\mathcal{R})$ is given by \cite{Messina2014,Messina2011PRA}:
\begin{widetext}
\begin{equation}
f_{\zeta}(\mathcal{R})=\left\{
\begin{array}{rcl}
\mathcal{P}_{-1}^{(\rm pw)} - \mathcal{R} \mathcal{P}_{-1}^{(\rm pw)} \mathcal{R}^{\dagger} + \mathcal{R} \mathcal{P}_{-1}^{(\rm ew)}  -  \mathcal{P}_{-1}^{(\rm ew)} \mathcal{R}^{\dagger}   & & \zeta = -1, \\ \\
\mathcal{P}_{1}^{(\rm pw)} - \mathcal{R}^{\dagger} \mathcal{P}_{1}^{(\rm pw)} \mathcal{R} + \mathcal{R}^{\dagger} \mathcal{P}_{1}^{(\rm ew)}  -  \mathcal{P}_{1}^{(\rm ew)} \mathcal{R}   & & \zeta = 1.
\end{array}
\right.
\label{auxiliary_function}
\end{equation}
\end{widetext}


According to Ref.~\cite{Messina2017PRB}, the periodicity along the $x$ axis makes it natural to replace the mode variable $k_x$ with $k_{xn}=k_x+n \frac{2\pi}{D}$, and $k_z$ becomes $k_{zn}=\sqrt{k_0^2- k_{xn}^2-k_y^2}$, where $n \in \mathbb{Z}$,  $k_{x}$ is in the first Brillouin zone $(-\frac{\pi}{D},\frac{\pi}{D})$, and $k_y \in \mathbb{R}$.

The reflection and transmission operators $\mathcal{R}^{(1)+}$, $\mathcal{R}^{(2)-}$, $\mathcal{T}^{(1)-}$ and $\mathcal{T}^{(2)-}$, needed for our calculations, are obtained from the operators $\mathcal{R}$ and $\mathcal{T}$ corresponding to the reference structure shown in Fig.~\ref{fig:shift_diagram} and whose derivation is detailed in the appendix. 
%
 \begin{figure} [htbp]
 \centerline {\includegraphics[width=0.5\textwidth]{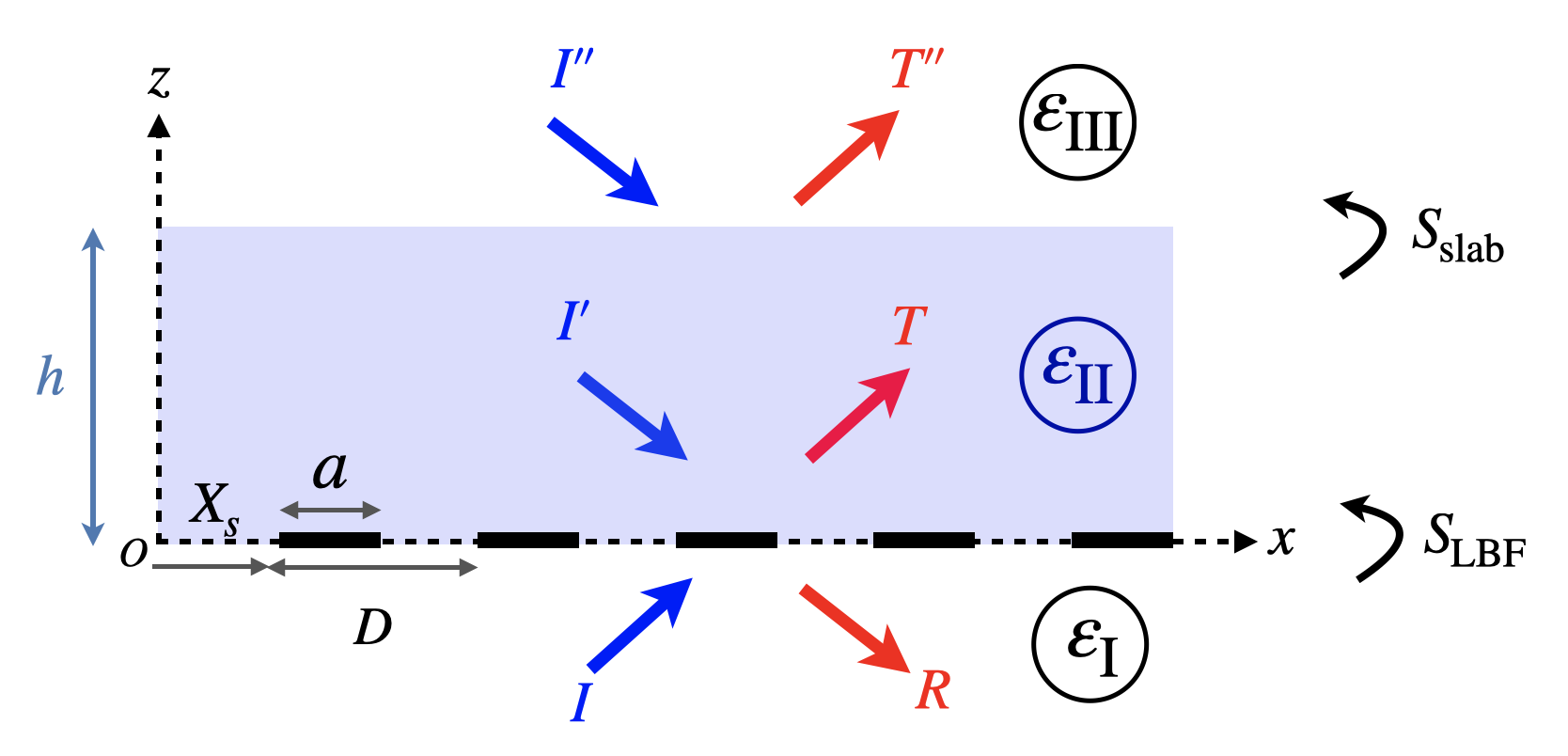}}
    \caption{Reference structure used in the computation of the different reflection and transmission operators.}
   \label{fig:shift_diagram}
 \end{figure}

Body 1 has no lateral shift, $i.e.$, $X_s=0$. Then $\mathcal{R}^{(1)+}$ and  $\mathcal{T}^{(1)-}$ at the interface $z=0$ and with no lateral shift $X_s=0$ (see Fig.~\ref{fig:two_gratings_schematic}) can be obtained directly from the known $\mathcal{R}^-$ and  $\mathcal{T}^+$ by using the following relations~\cite{Messina2017PRB,Luo2023apl_NFRHT_grating}: 
\begin{equation}
\begin{split}
&\left \langle p,{\textbf{k}},n|\mathcal{R}^{(1)+}(\omega)|p',{\textbf{k}}',n' \right \rangle \\ &=  \left\{
\begin{array}{rcl}
\begin{aligned}
&\left \langle p,{\textbf{k}},n|\mathcal{R}^{-}(\omega)|p,{\textbf{k}}',n' \right\rangle & p=p'  \\
&-\left \langle p,{\textbf{k}},n|\mathcal{R}^{-}(\omega)|p',{\textbf{k}}',n' \right\rangle & p\neq p' ,  
\end{aligned}
\end{array}
\right.
\end{split}
\label{R1+}
\end{equation}
%
and
%
\begin{equation}
\begin{split}
& \left \langle p,{\textbf{k}},n|\mathcal{T}^{(1)-}(\omega)|p',{\textbf{k}}',n' \right\rangle \\& =\left\{
\begin{array}{rcl}
\begin{aligned}
& \left \langle p,{\textbf{k}},n|\mathcal{T}^{+}(\omega)|p,{\textbf{k}}',n' \right\rangle, p=p' 
\\
&  -\left \langle p,{\textbf{k}},n|\mathcal{T}^{+}(\omega)|p',{\textbf{k}}',n' \right\rangle, p\neq p' .  
\end{aligned}
\end{array}
\right.
\end{split}
\label{nondiagnoal}
\end{equation}

As for body 2, there is, not only, a lateral shift $X_s$ but an interface translation to $z=d$ (see Fig.~\ref{fig:two_gratings_schematic}).  In order to take into account these two kinds of translation and obtain $\mathcal{R}^{(2)-}$ and  $\mathcal{T}^{(2)-}$ one can consider two methods. The first one consists in directly applying appropriate translation operators to the scattering operators. This approach has been discussed, in detail, in a previous work \cite{Messina2011PRA} from which we take the results giving $\mathcal{R}^{(2)-}$ and  $\mathcal{T}^{(2)-}$ in terms of $\mathcal{R}^{-}$ and  $\mathcal{T}^{-}$:
\begin{equation}
\left\{
\begin{array}{rcl}
\begin{aligned}
 &\left \langle p,{\textbf{k}},n|\mathcal{R}^{(2)-}(\omega)|p',{\textbf{k}}',n' \right\rangle 
 \\&=e^{i\left(k_{xn'}' - k_{xn}\right)X_s}  e^{i\left(k_{zn}^{}+k_{zn'}'\right)d}	\left \langle p,{\textbf{k}},n|\mathcal{R}^{-}(\omega)|p',{\textbf{k}}',n' \right\rangle,
\\&\left \langle p,{\textbf{k}},n|\mathcal{T}^{(2)-}(\omega)|p',{\textbf{k}}',n' \right\rangle 
 \\&=e^{i\left(k_{xn'}' - k_{xn}\right)X_s}  e^{i\left(k_{zn}^{}-k_{zn'}'\right)d}	\left \langle p,{\textbf{k}},n|\mathcal{T}^{-}(\omega)|p',{\textbf{k}}',n' \right\rangle.
\end{aligned}
\end{array}
\right.
\label{R2_T2_}
\end{equation}

In the first method, when using the Eq.~(\ref{R2_T2_}), $\mathcal{R}^{-}$ and $\mathcal{T}^{-}$ are calculated by fixing $X_s=0$, since the effect of lateral shift is already included by the factor phase $e^{i\left(k_{xn'}' - k_{xn}\right)X_s}$.

The second method, consists in incorporating the effect of the lateral shift directly in the computation of the scattering operators $\mathcal{R}^{-}$ and  $\mathcal{T}^{-}$ with the FMM-LBF (as given in the appendix), and then only applying the normal translation operators $e^{i\left(k_{zn}^{}+k_{zn'}'\right)d}$ and $e^{i\left(k_{zn}^{}-k_{zn'}'\right)d}$ (as previously) in the $z$ direction.    

\begin{figure} [htbp]
\centerline {\includegraphics[width=0.45\textwidth]{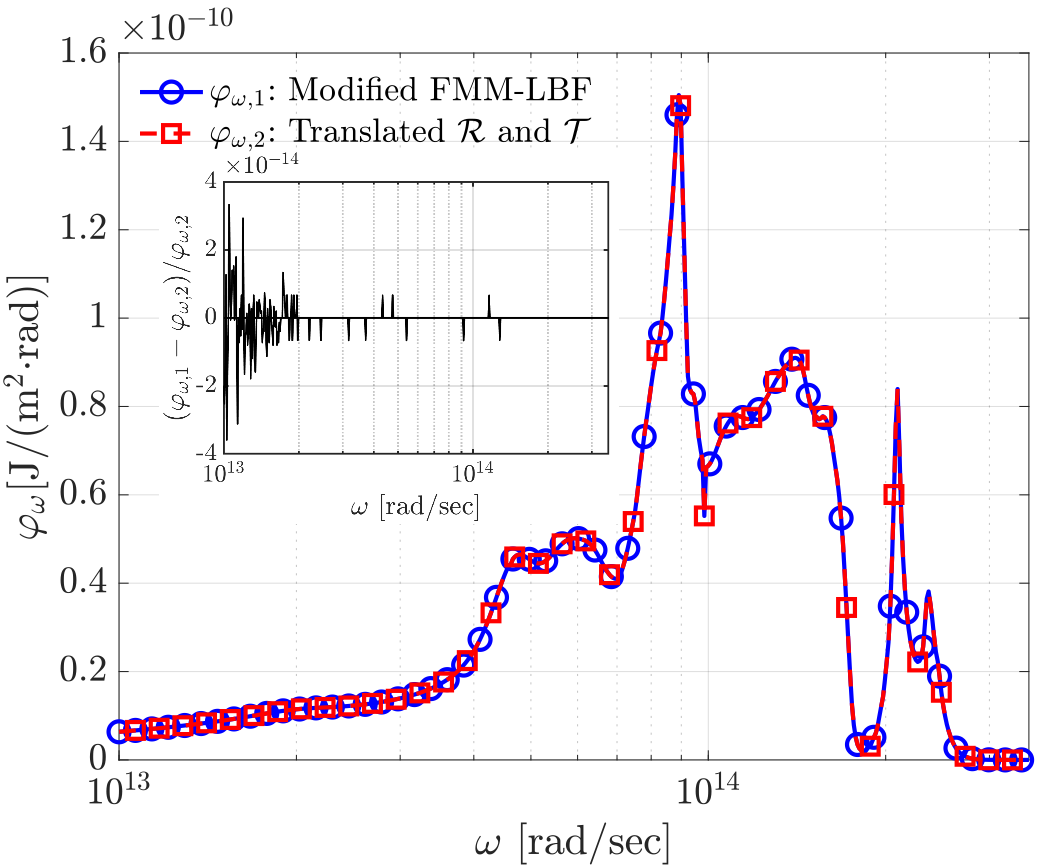}} 
     \caption{The radiative heat flux spectrum of the configuration shown in Fig.~\ref{fig:two_gratings_schematic} using SiO$_2$ slabs (whose optical data are taken from Ref.\cite{Palik}), $\mu=0.3$ eV, a lateral shift $X_s=0.5\mu$m and a grating period $D=2\mu$m at separation distance $d=100$ nm, with $T_1=290$K and $T_2=310$K, obtained by the two different approaches : (1) the modified FMM-LBF and (2) the translated scattering operators}
   \label{fig:modified_LBFs}
\end{figure}

It is important to stress that these two approaches are completely equivalent. A numerical verification of this is shown in Fig.~\ref{fig:modified_LBFs} where we compute the radiative heat flux spectrum for the structure shown in Fig.~\ref{fig:two_gratings_schematic}, using the two methods (the different parameters are given in the caption). As can be seen from the figure, there is a very good agreement between the two approaches and a closer look at the data reveals that the results are identical to our working precision {[See the inset of the figure for the near 0 relative ratio of heat flux $(\varphi_{\omega,1}-\varphi_{\omega,2})/\varphi_{\omega,2}$ between the two methods]}. Therefore, either of these methods can be used in the investigations to come.     

\section{Results and discussion}
\label{results_discussion}
In this section, we use the  theoretical model outlined above to investigate the NFRHT between the slabs of Fig.~\ref{fig:two_gratings_schematic}. Parametric investigations of influencing factors on NFRHT will be performed and a particular interest will be given to the asymptotic regimes for the lateral shift effect. In the following, bodies 1 and 2 are maintained at $T_1=290$K and $T_2=310$K, respectively. {As for the truncation order $N$ used in the FMM-LBF, we notice that the convergence of truncation order $N$ used in the FMM-LBF is independent of the homogeneous media surrounding graphene and the conductivity itself. More precisely, this $N$ convergence depends on the ratio between the wavelength and the period of the grating. In our computations, given the interval of wavelengths considered, we found that $N = 30$ is enough to ensure the convergent results for the radiative heat flux (the relative error <1\%).}
 The substrate thickness $h$ is fixed at 20 nm.  



\subsection{Parametric investigation of influencing factors on lateral-shift-induced NFRHT: grating period, chemical potential and filling fraction}
\label{Sec_Para_inves}

\begin{figure*} [htbp]
\centerline {\includegraphics[width=0.9\textwidth]{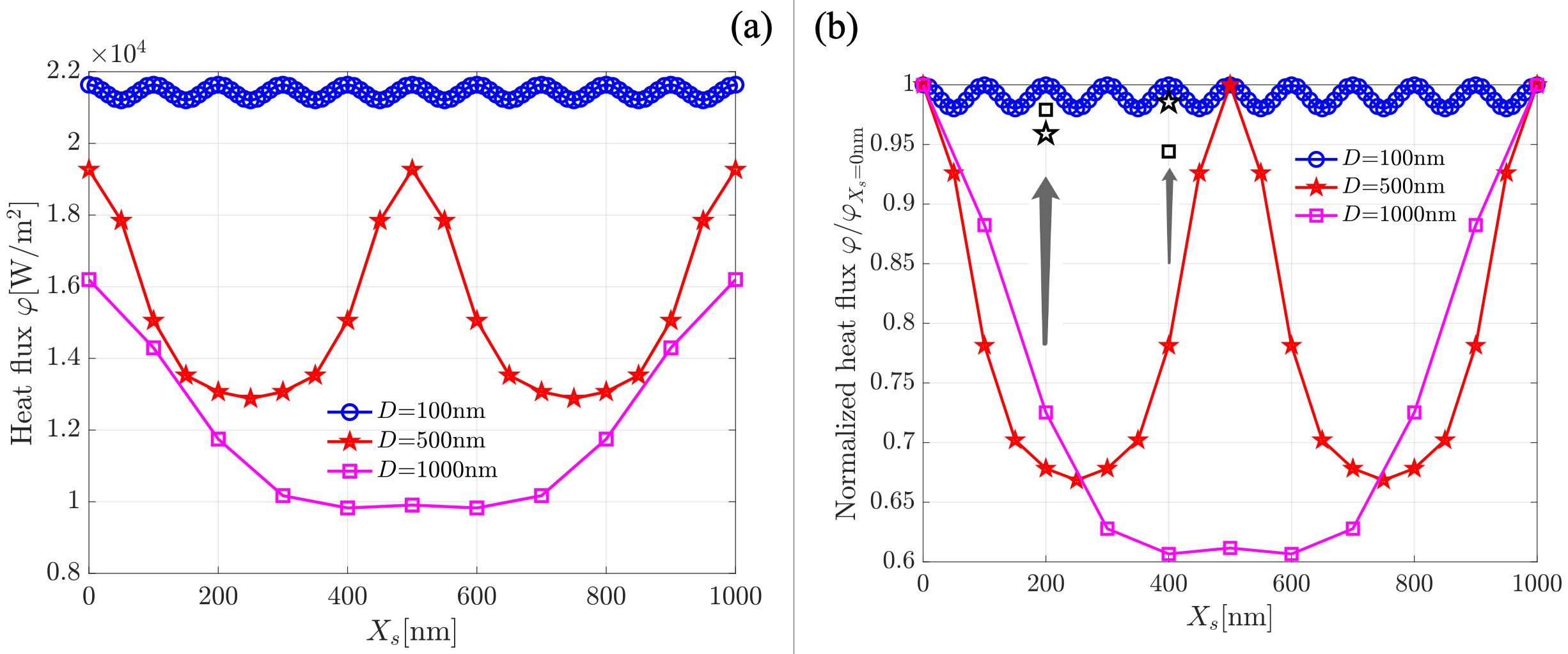}}
     \caption{Dependence of heat flux $\varphi$ and normalized heat flux $\varphi/\varphi(X_s=0\text{nm})$ on the relative lateral shift $X_s$: (a) the absolute heat flux $\varphi$ and (b) ratio of the heat flux $\varphi$ to the heat flux at $X_s=0$nm. Three different periods are considered, $D=100$nm, 500nm and 1000nm.}
   \label{fig:period_effect}
\end{figure*}

We consider three different influencing factors that may affect the effect of relative lateral shift between the two gratings on its heat transfer: the grating period $D$, the chemical potential $\mu$ of graphene and the filling fraction $f$. We start the parametric investigation with the grating period $D$. In Fig.~\ref{fig:period_effect} (a), we show the dependence of heat flux $\varphi$ on the relative lateral shift $X_s$. Three different periods are considered, $D=100$nm, 500nm and 1000nm, respectively. The other parameters are $f=0.5$, $d=100$nm, and $\mu=0.5$eV. 

When shifting one of the grating, we can observe a periodic oscillation of the radiative heat flux with the same spatial period as the grating itself. As shown in Fig.~\ref{fig:period_effect} (a), for a fixed lateral shift $X_s$, the radiative heat flux decreases by about 30\% when the period $D$ increases from 100nm to 1000nm. Additionally, the oscillation amplitude of the radiative heat flux induced by the lateral shift increases significantly with increasing period $D$. To show the change of the oscillation amplitude corresponding to different grating periods, we normalize the radiative heat flux $\varphi(X_s)$ by $\varphi(X_s=0\text{nm})$ (\textit{i.e.} the one without a lateral shift). The result is shown in Fig.~\ref{fig:period_effect} (b). Lateral shift affects the heat transfer significantly. For the case where $D=1000$nm, the lateral shift can even result in 40\% reduction of the heat flux. While for the case where $D=100$nm, the lateral shift can reduce it by only few percent, which is negligible compared to that for the case of a large spatial period ($D=1000$nm and 500nm). For a fixed separation distance $d$, the scattering details become more important when increasing the grating period $D$ (and thus the ratio $D/d$), causing the nanostructures "see" each other in more detail. In such a situation, a {more accurate} is needed to account for the complexity of the nanostructures and be able to compare with experiments while the effective medium theory becomes completely invalid. 

\begin{figure} [htbp]
\centerline {\includegraphics[width=0.5\textwidth]{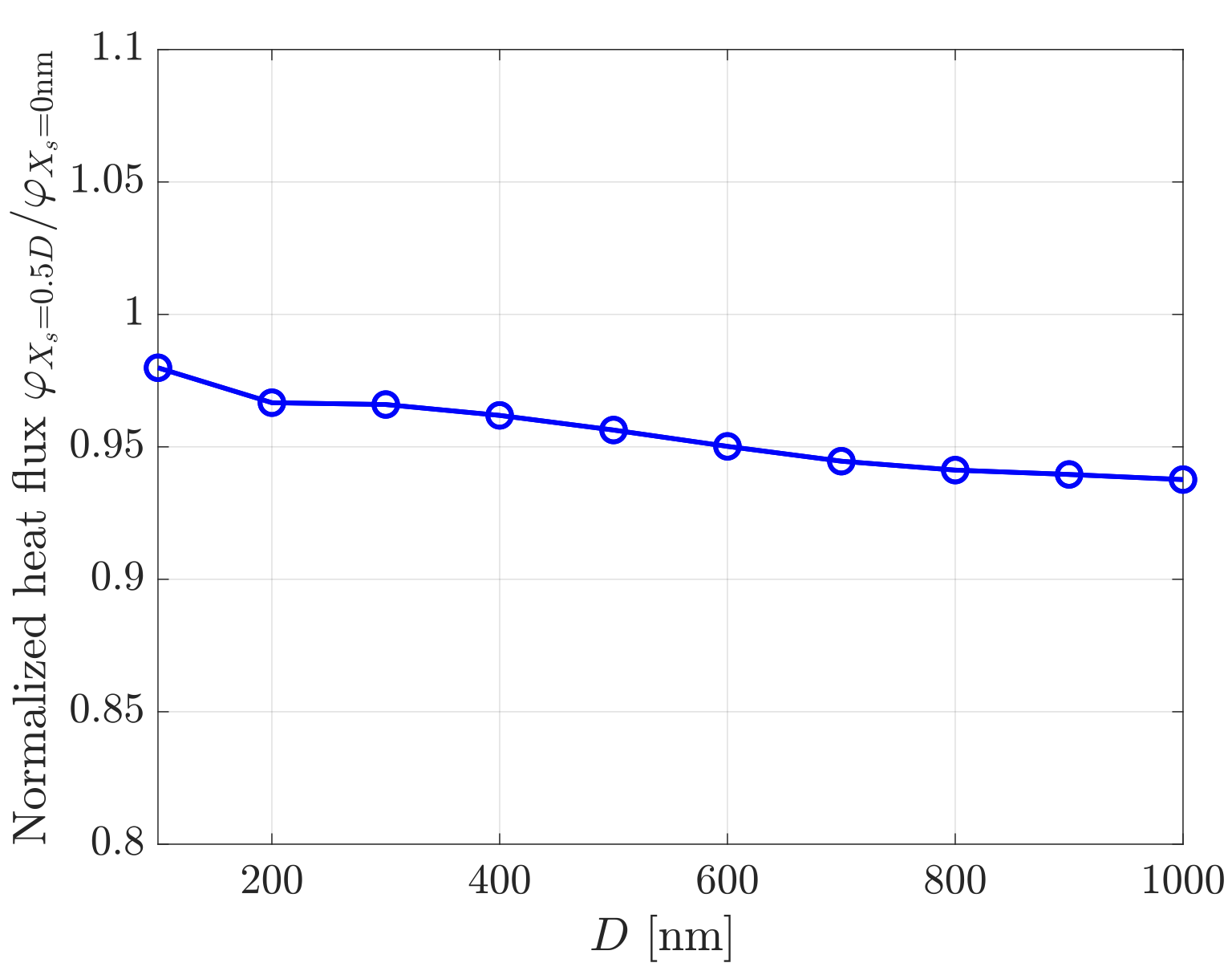}}
     \caption{Dependence of the ratio $\varphi_{X_s=0.5D}^{}/\varphi_{X_s=0{\rm nm}}^{}$ on the period $D$. Here: $d/D= 1$, $f=0.5$, $\mu=0.5$eV. }
   \label{fig:d_D_1}
\end{figure}

In our recent paper \cite{Luo2023Casimir_gg}, when discussing the non-additivity of the Casimir force (a kind of geometric effect) between two aligned graphene grating coated finite-size fused silica slabs, (the same configuration as in this work but with no lateral shift), we introduced a dimensionless parameter $d/D$ and proved its relevance to such a kind of geometric effect. The effect of the relative lateral shift on the NFRHT is also geometry related, therefore, we want to know if the dimensionless parameter $d/D$ is still pertinent in this case. The three curves in Fig.~\ref{fig:period_effect} (b) correspond to a fixed separation distance $d=100$nm, where the $d/D=1$, 0.2 and 0.1 for the cases $D=100$nm, 500nm and 1000nm, respectively. To verify the relevance of $d/D$ to the lateral shift effect, we performed additional calculations using the couples ($d=500$nm, $D=500$nm) and ($d=1000$nm, $D=1000$nm) to in order to maintain $d/D=1$ as for the case ($d=100$nm, $D=100$nm) already shown. Two lateral shift distances $X_s=200$nm and 400nm are considered. The results are added in Fig.~\ref{fig:period_effect} (b) and shown as the star and square symbols in black for period $D=500$nm and 1000nm respectively. We see that they both move close to the $d=100$nm and $D=100$nm curve, confirming the relevance of the geometric factor $d/D$ and hence of the geometric nature of the effect.  {To further verify the relevance of $d/D$ to the lateral shift effect, the dependence of the amplitude ratio $\varphi_{Xs = 0.5D}^{}/\varphi_{Xs = 0}^{}$ on the period $D$ is shown in Fig.~\ref{fig:d_D_1} where the $d/D$ is fixed at 1. We observe that the ratio $\varphi_{Xs = 0.5D}^{}/\varphi_{Xs = 0}^{}$ is close to 1 for the considered periods (100nm < $D$ < 1000nm). That is, maintaining $d/D = 1$ will bring a weak lateral shift effect compared to the significant lateral shift effect for the configurations (d/D = 0.2 and 0.1) in Fig.~\ref{fig:period_effect} (b), which further confirms the relevance of the geometric factor $d/D$ to the lateral shift effect.}

\begin{figure} [htbp]
\centerline {\includegraphics[width=0.5\textwidth]{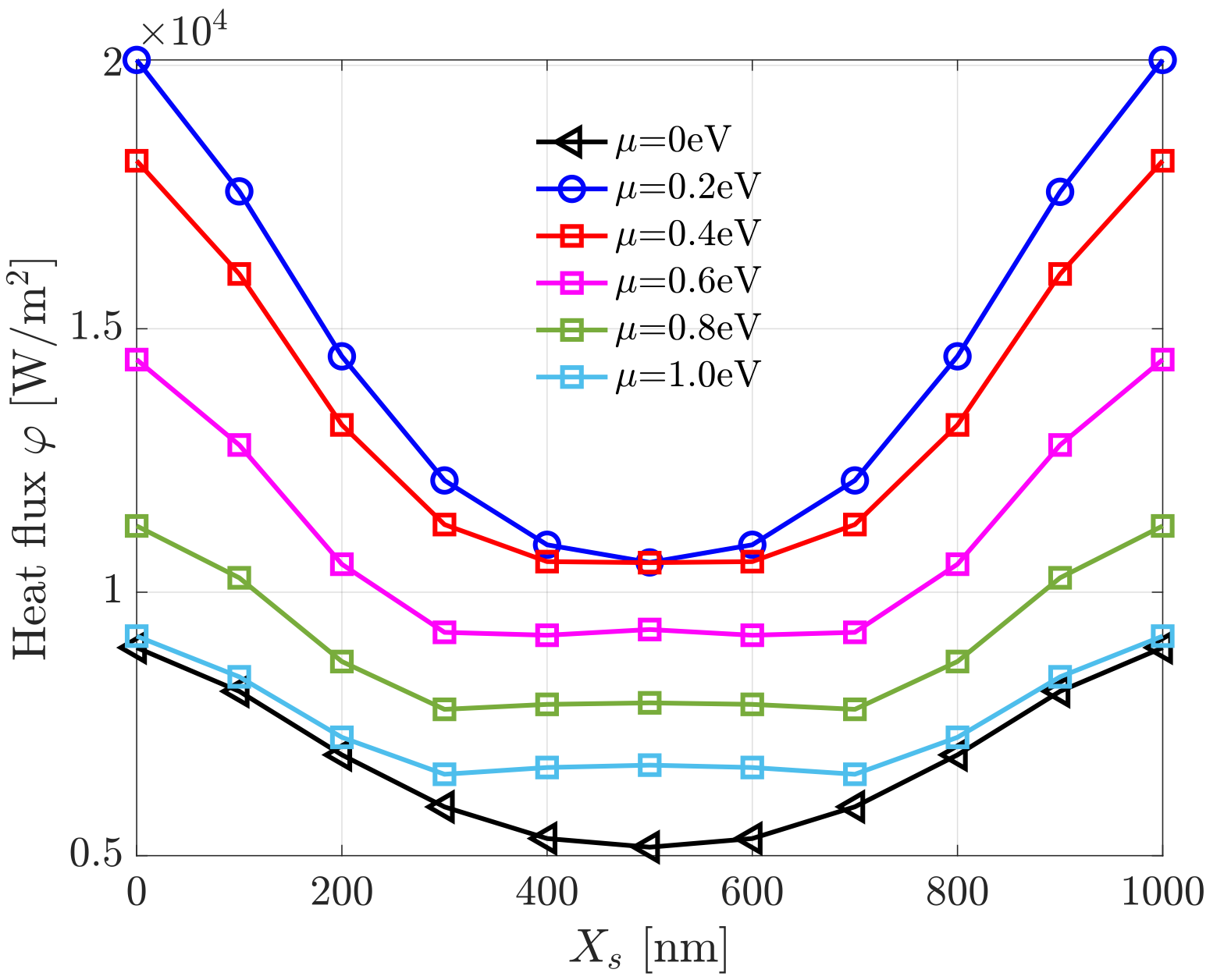}}
     \caption{Dependence of radiative heat flux $\varphi$ on the lateral shift $X_s$. The chemical potential $\mu=0$eV, 0.2eV, 0.4eV, 0.6eV, 0.8eV and 1.0eV. The period $D=1000$nm, filling fraction $f=0.5$, separation distance $d=100$nm.}
   \label{fig:chemical_potential_effect}
\end{figure}

\begin{figure} [htbp]
\centerline {\includegraphics[width=0.5\textwidth]{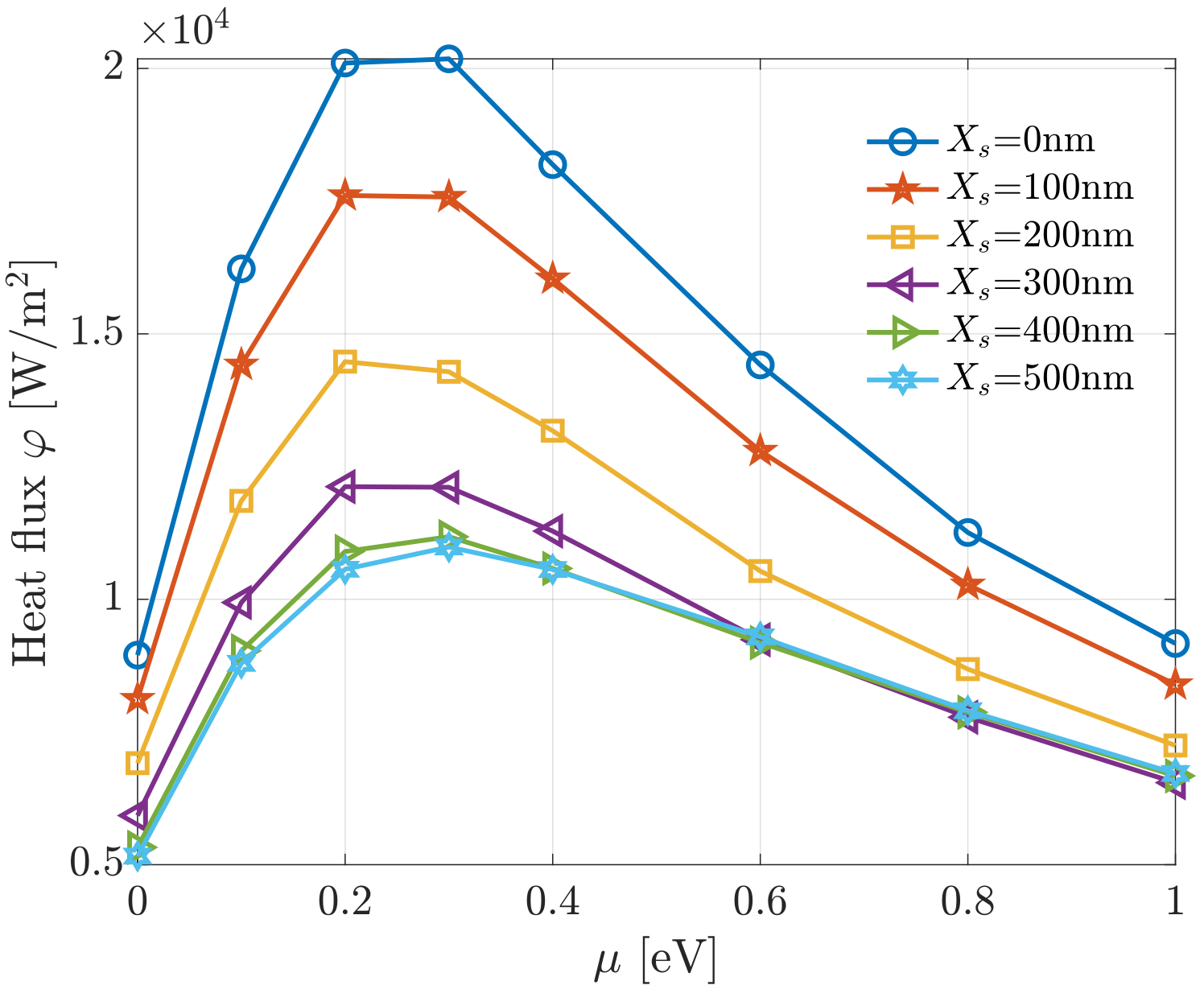}}
     \caption{Dependence of radiative heat flux $\varphi$ on the chemical potential $\mu$ of graphene. The lateral shift $X_s=0$nm, 100nm, 200nm, 300nm, 400nm and 500nm, period $D=1000$nm, filling fraction $f=0.5$, separation distance $d=100$nm.}
   \label{fig:chemical_potential_optimal}
\end{figure}

We then study the effect of the chemical potential $\mu$ of graphene on the NFRHT. We show the dependence of heat flux $\varphi$ on the relative lateral shift $X_s$ in Fig.~\ref{fig:chemical_potential_effect}, considering different chemical potential values $\mu=0$eV, 0.2eV, 0.4eV, 0.6eV, 0.8eV and 1.0eV with $D=1000$nm, $f=0.5$ and $d=100$nm. The upper body 2 is laterally shifted in the range of one full grating spatial period. At different Fermi levels, the lateral shift effect on the NFRHT changes slightly. When the chemical potential is not too small ($e.g.,\mu>0.2$eV), when laterally shifting body 2, the radiative heat flux decreases to reach a valley, then keeps constant for a range of lateral shifts (forming a plateau), and finally increases to the same value as that without lateral shift. While for the chemical potential $\mu=0$eV and 0.2eV, radiative heat flux will decrease to reach the valley and then increase directly and go back to that of no lateral shift. Fig.~\ref{fig:chemical_potential_effect} also indirectly shows that the radiative heat flux is not monotonic with the chemical potential (this behavior will be studied in detail in Fig.~\ref{fig:chemical_potential_optimal}). A natural question is whether we can obtain an optimal radiative heat flux by choosing an appropriate chemical potential $\mu$ for graphene gratings.

To answer the above question, we calculated the dependence of the radiative heat flux $\varphi$ on the chemical potential $\mu$ for different lateral shifts $X_s= 0$nm, 100nm, 200nm, 300nm, 400nm and 500nm, for $D = 1000$nm, $f = 0.5$ and $d = 100$nm. The results are reported in Fig.~\ref{fig:chemical_potential_optimal} where we see that when increasing the chemical potential, the radiative heat flux increases at first to its peak and then decreases gradually for all considered lateral shift $X_s$ curves. For the considered separation $d$ and grating period $D$, the radiative heat flux appears to be optimal for $\mu $ at around $0.25$eV. 

\begin{figure} [htbp]
\centerline {\includegraphics[width=0.5\textwidth]{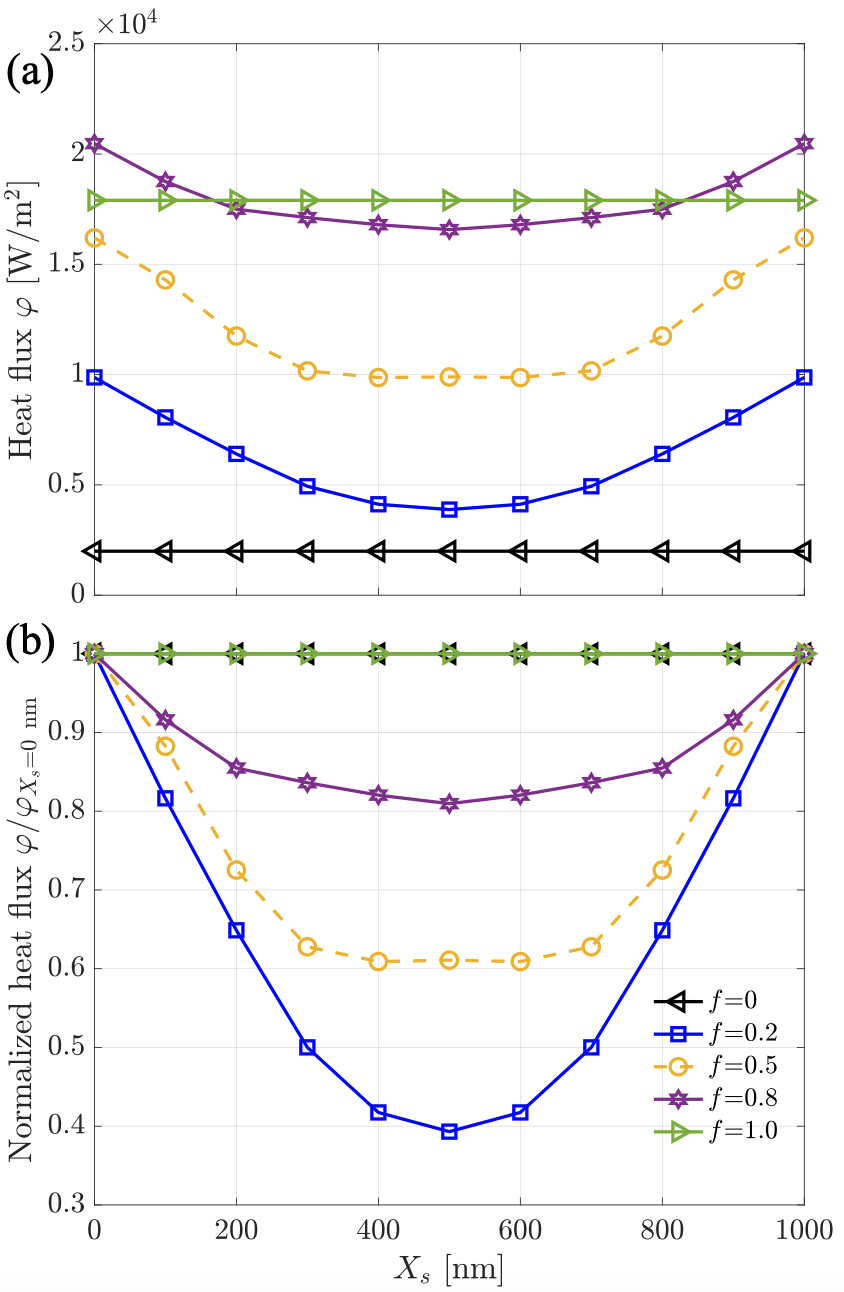}}
     \caption{Dependence of heat flux $\varphi$ and normalized heat flux $\varphi/\varphi(X_s=0\text{nm})$ on the relative lateral shift $X_s$: (a)the absolute heat flux  $\varphi$ and (b) ratio of the heat flux  $\varphi$  to the heat flux at $X_s = 0$nm. Five different filling fractions are considered, $f$ = 0, 0.2, 0.5, 0.8 and 1.0, respectively. The separation distance $d=100$nm, grating period $D=1000$nm, chemical potential $\mu=0.5$eV.}
   \label{fig:filling_fraction_effect}
\end{figure}

Additionally, we study the effect of the filling fraction $f$ on the NFRHT. We show the dependence of heat flux on the filling fraction in Fig.~\ref{fig:filling_fraction_effect} (a) where five different filling fractions are considered, $f$ = 0, 0.2, 0.5, 0.8 and 1.0 with $d=1000$nm and $D=1000$nm. For $f$ = 0 and 1.0, the configurations reduce to the bare/coated slab cases. As expected, the heat flux for these two cases is independent on the lateral shift. At a large separations (\textit{e.g.}, $d=1000$nm), adding a graphene sheet coating on the fused silica substrate (right triangle green line) will significantly enhance the radiative heat flux as compared to the bare fused silica substrate (left triangle black line), which is consistent with the observations in references \cite{Luo2023apl_NFRHT_grating,Svetovoy2012prb}. At short separations and without considering any lateral shift, as reported in our recent work \cite{Luo2023apl_NFRHT_grating}, patterning graphene from a sheet to a grating, can significantly enhance the heat flux. {For the graphene grating coating case (\textit{e.g.}, $f=0.8$), there are more accessible high-$k$ modes, while for the graphene sheet configuration ($f=1.0$), the accessible wavevector region is relatively smaller than that of the graphene grating one. Consequently, the slabs coated with a graphene grating can facilitate a greater transfer of energy compared to the slabs coated with a graphene sheet.} However, at large separations, the patterning method does not always work to enhance heat flux. As shown in Fig.~\ref{fig:filling_fraction_effect} (a), by comparing the curves for $f=0.2$, 0.5 and 0.8 to the curve for $f=1.0$, we can clearly see that, apart from the enhancement seen for $f=0.8$, the patterning method fails to enhance the heat flux. Moreover, when laterally shifting the upper body 2 in the case $f=0.8$, the heat flux will become less than that of the graphene sheet coating case ($f=1.0$) (for $X_s \in [200,800]$nm).

In Fig.~\ref{fig:filling_fraction_effect} (b), we show the dependence of the ratio $\varphi(X_s)/\varphi(X_s=0)$ on the filling fraction $f$. As expected, it is constant an equal to 1 for both $f=0$ and 1. For the other filling fractions, this ratio decreases from 1 to a valley value and then increases back to 1. For a lateral shift about 500nm, the heat flux can even experience up to 60\% reduction when $f=0.2$. {In general, the minimal heat flux is at the half period lateral shift ($X_s=0.5D$). However, it is not always the case. For example, as shown in Fig.~\ref{fig:period_effect} (b), the minimal heat flux is at $X_s=400$nm and $X_s=600$nm rather than at $X_s=0.5D$. According to Fig.~\ref{fig:period_effect}, Fig.~\ref{fig:chemical_potential_effect} and Fig.~\ref{fig:filling_fraction_effect}, the lateral shift for the minimum heat flux is relevant to the grating period $D$, chemical potential $\mu$ and the filling fraction $f$.} Considering that both the scale of the lateral shift  and the geometry, considered here, are realistic and experimentally accessible, the lateral-displacement-sensitive radiative heat transfer might have potential for thermal logic gates applications.

\subsection{Asymptotic regimes for the lateral shift effect on NFRHT}
In this section, we try to find an asymptotic regime map for the lateral shift effect on NFRHT between substrate-supported graphene gratings to quickly tell the existence of the lateral shift effect or not. Considering that the relevance of the dimensionless geometric factor $d/D$ for the lateral shift effect has been confirmed in Section \ref{Sec_Para_inves}, we may take advantage of it to propose the asymptotic regime map. That is, as the filling fraction $f$ approaches 0 or 1, the lateral shift effect disappears. Hereinafter, we will use the ratio of heat flux with a half period shift $\varphi (X_s=0.5D)$ to that without a shift $\varphi (X_s=0)$  to evaluate the lateral shift effect on heat flux.

\begin{figure} [htbp]
\centerline {\includegraphics[width=0.5\textwidth]{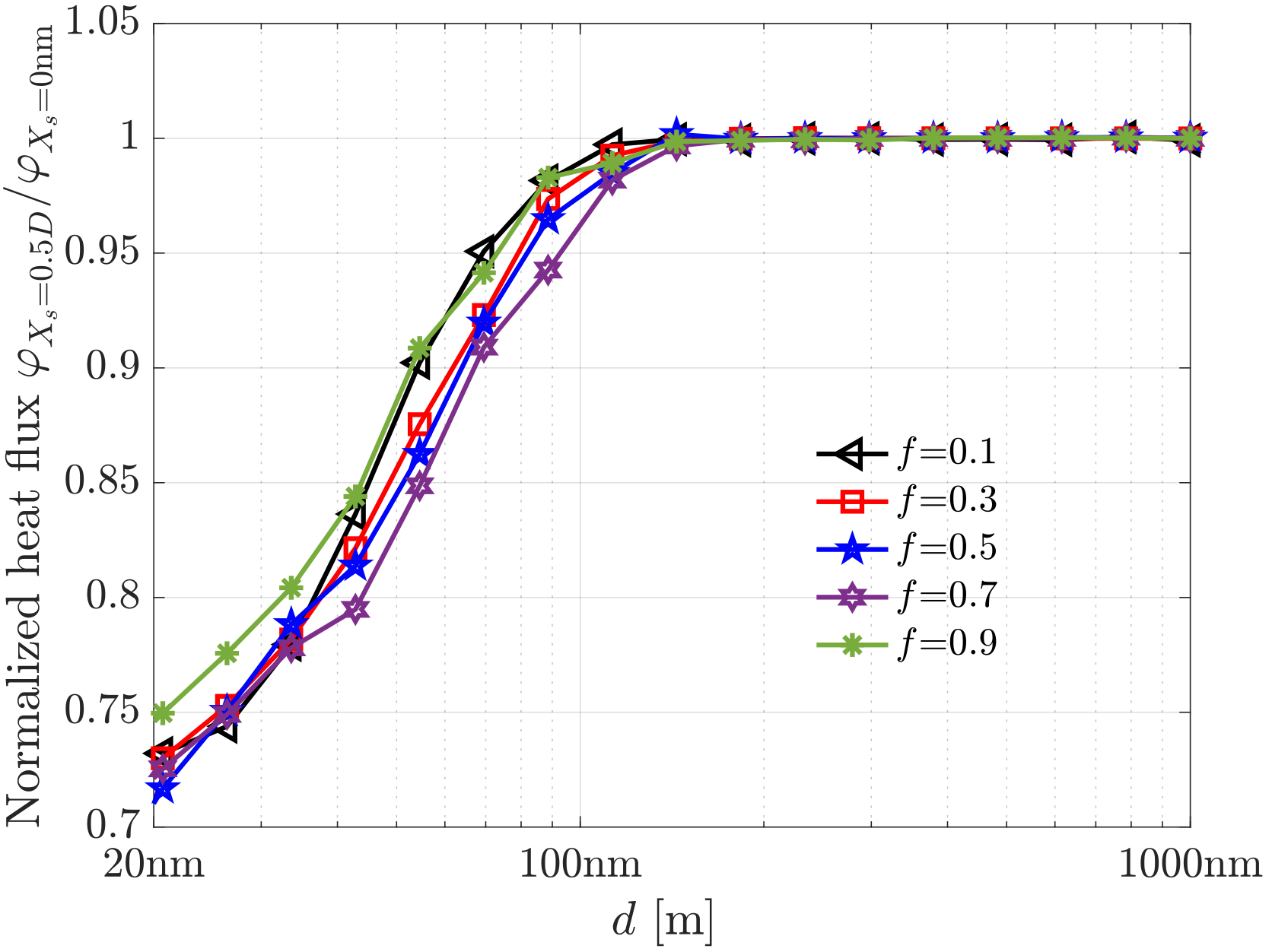}}
     \caption{Dependence of normalized heat flux $\varphi_{X_s=0.5D}^{}/\varphi_{X_s=0\text{nm}}^{}$ on separation distance $d$. Five different filling fractions are considered, $f$ = 0.1, 0.3, 0.5, 0.7 and 0.9, respectively. The grating period $D=100$nm, chemical potential $\mu=0.5$eV.}
   \label{fig:regime_a}
\end{figure}

We will, first, check whether the filling fraction affects the asymptotic regime. The dependence of the ratio $\varphi (X_s=0.5D)/\varphi (X_s=0)$ on the separation distance $d$ is shown in Fig.~\ref{fig:regime_a} where five different filling fractions are considered : $f$ = 0.1, 0.3, 0.5, 0.7 and 0.9. The grating period is $D=100$nm and the chemical potential is $\mu=0.5$eV. We can see a clear dependance of the ratio $\varphi (X_s=0.5D)/\varphi (X_s=0)$ on $d$ with two distinct regime zones:  (i) the first one for $d \le 200$nm where the ratio varies quickly and (ii) the second one for $d \ge 200$nm where the ratio is almost constant. Moreover, we remark that even for small period configurations (\textit{i.e.}, 100nm), the lateral shift can still have a wide range modulation of the heat flux, with a maximum reduction of 30\% (in the first zone).
%

\begin{figure*} [htbp]
\centerline {\includegraphics[width=0.7\textwidth]{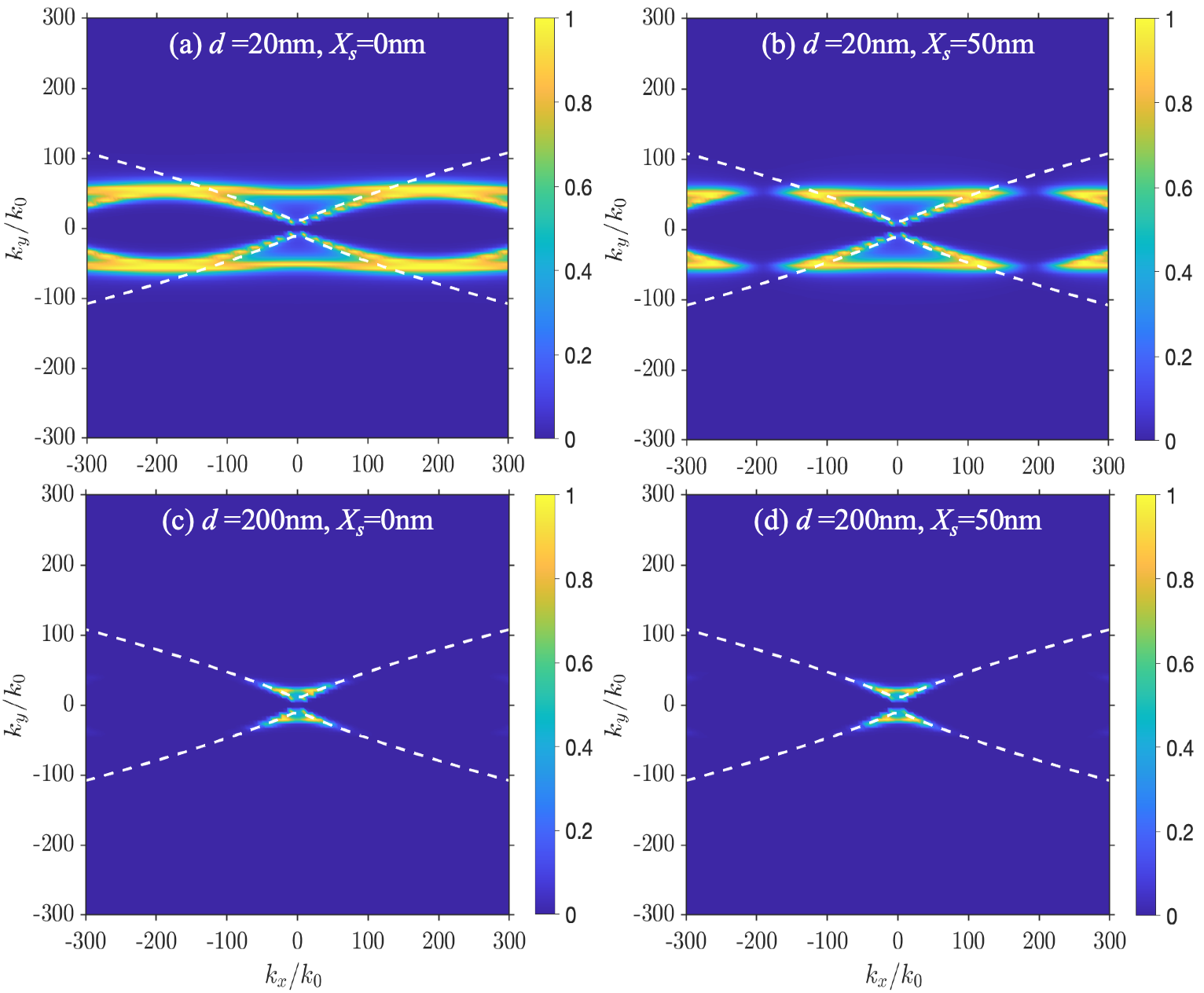}}
     \caption{Energy transmission coefficient for a fused silica substrate coated with a graphene grating: (a) separation $d=20$nm and lateral shift $X_s=0$nm, (b) separation $d=20$nm and lateral shift $X_s=50$nm, (c) separation $d=200$nm and lateral shift $X_s=0$nm, and (d) separation $d=200$nm and lateral shift $X_s=50$nm. The grating period $D=100$nm, chemical potential $\mu=0.5$eV, filling fraction $f$=0.5.}
   \label{fig:regime_a_Ener_Trans}
\end{figure*}

 Beyond the separation point $d=200$nm, the lateral shift effect becomes less important. In addition, such a critical separation point does not change with the filling fraction $f$. To understand why the lateral shift effect on heat transfer changes with the separation $d$, we show the energy transmission coefficient Tr($\mathcal{O}$) in the ($k_x/k_0$, $k_y/k_0$) plane at the angular frequency $\omega=5\times 10^{13}$rad/s (where the contribution of graphene to the NFRHT spectrum is important) for a slab coated with a graphene grating in Fig.~\ref{fig:regime_a_Ener_Trans}. Tr($\mathcal{O}$) is the sum over the two polarizations of the photon tunneling probabilities, which is usually applied to analyze the mechanisms behind the NFRHT \cite{Greffet2012prb_trans_def,Liu2014apl_trans_def,Luo2023apl_NFRHT_grating}. The transmission coefficient operator $\mathcal{O}$ is defined by Eq.~(\ref{O_operator}). In Fig.~\ref{fig:regime_a_Ener_Trans}, four configurations are considered: (a) $d=20$nm and $X_s=0$nm, (b) $d=20$nm and $X_s=50$nm, (c) $d=200$nm and t $X_s=0$nm, and (d)  $d=200$nm and $X_s=50$nm, together with $D=100$nm, $\mu=0.5$eV and $f$=0.5. The dotted curves represent the dispersion relations obtained from the poles of the reflection coefficient \cite{Zhou2022langmuir,Zhou2022prm}. 
%
%

For the configurations with $d=200$nm, the topology of the accessible modes does not change with the lateral shift, as shown in Fig.~\ref{fig:regime_a_Ener_Trans} (c) and (d). However, for $d=20$nm, the topology of the accessible modes changes slightly, as shown in Fig.~\ref{fig:regime_a_Ener_Trans} (a) and (b). There are more accessible high-$k$ modes for the case with no lateral shift than that for the case with a half-period lateral shift, which accounts for the fact that the radiative heat flux decreases significantly as the upper body 2 moves from the aligned situation to the misaligned one. Whether the lateral shift exists or not, the supported surface plasmon polariton is alway the hyperbolic one. Compared to the transition from circular one to the hyperbolic one by patterning the graphene sheet into a grating, the lateral shift will not induce a critical topology transition, but only slightly affect the accessible range of high-$k$ modes.

\begin{figure} [htbp]
\centerline {\includegraphics[width=0.5\textwidth]{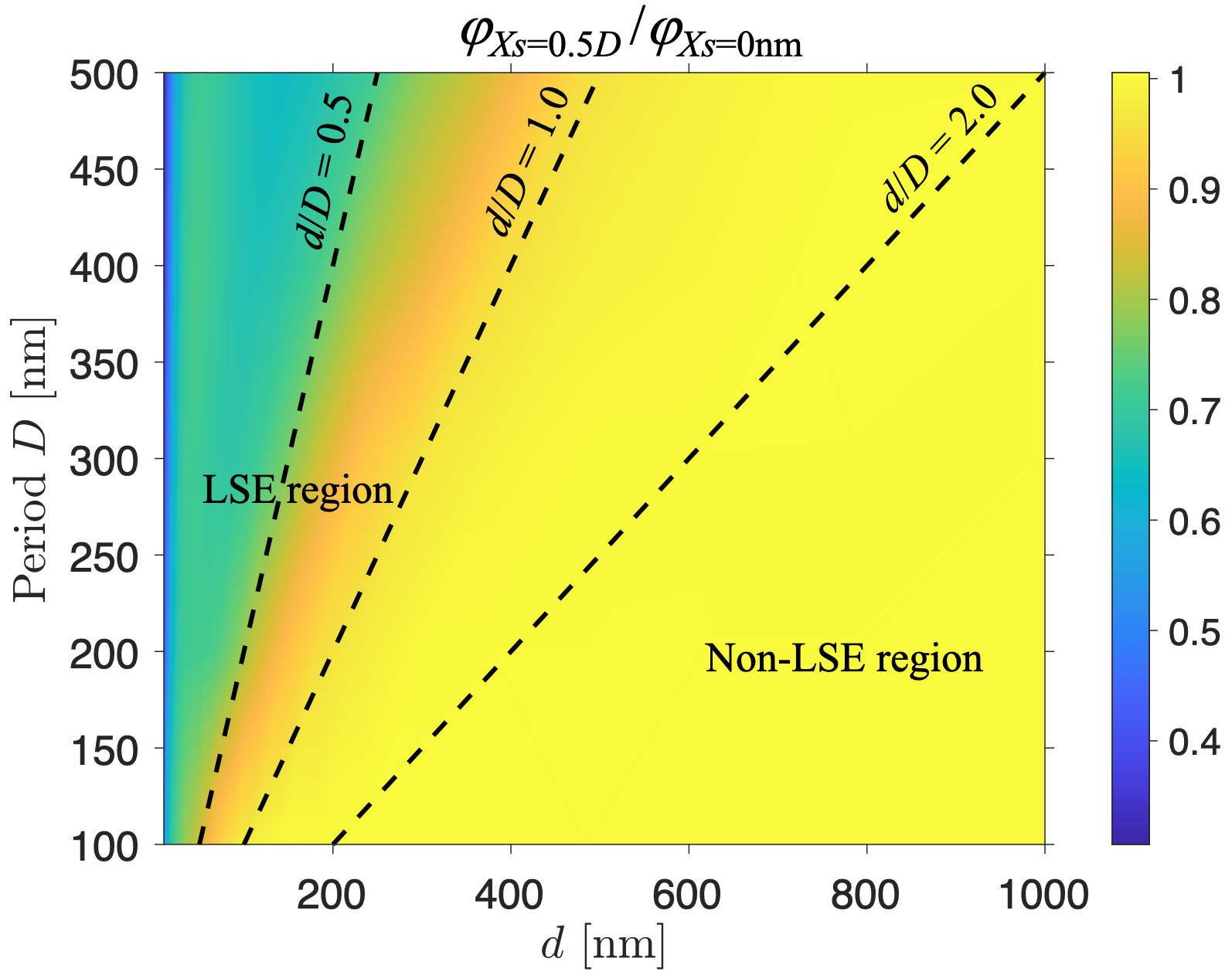}}
     \caption{Dependence of the ratio $\varphi_{X_s=0.5D}^{}/\varphi_{X_s=0\text{nm}}^{}$ on separation distance $d$ and grating period $D$. The filling fraction $f$ = 0.5, chemical potential $\mu=0.5$eV. The dotted lines (geometric factor $d/D=0.5$, 1.0 and 2.0, respectively) are added for reference.}
   \label{fig:regime_b}
\end{figure}

Now, we investigate and determine the asymptotic regime map for the lateral shift effect by using the geometric factor $d/D$. For that, we take the typical filling fraction $f=0.5$ and chemical potential $\mu=0.5$eV. In Fig.~\ref{fig:regime_b} we show the ratio $\varphi_{X_s=0.5D}^{}/\varphi_{X_s=0\text{nm}}^{}$ in the ($d,D$) plane where the black dotted lines corresponding to the lines representing the geometric factor $d/D=0.5$, 1.0 and 2.0 are added for reference.

%

For a given grating period $D$, as the separation $d$ becomes large enough, the ratio $\varphi_{X_s=0.5D}^{}/\varphi_{X_s=0\text{nm}}^{}$ approaches 1 and the scattering details of the graphene grating do not change with the relative lateral shift $X_s$ anymore. That is, the two graphene gratings see each other as an effective whole rather than in detail, and thus the lateral shift effect on heat transfer becomes less important. 

\begin{figure} [htbp]
\centerline {\includegraphics[width=0.5\textwidth]{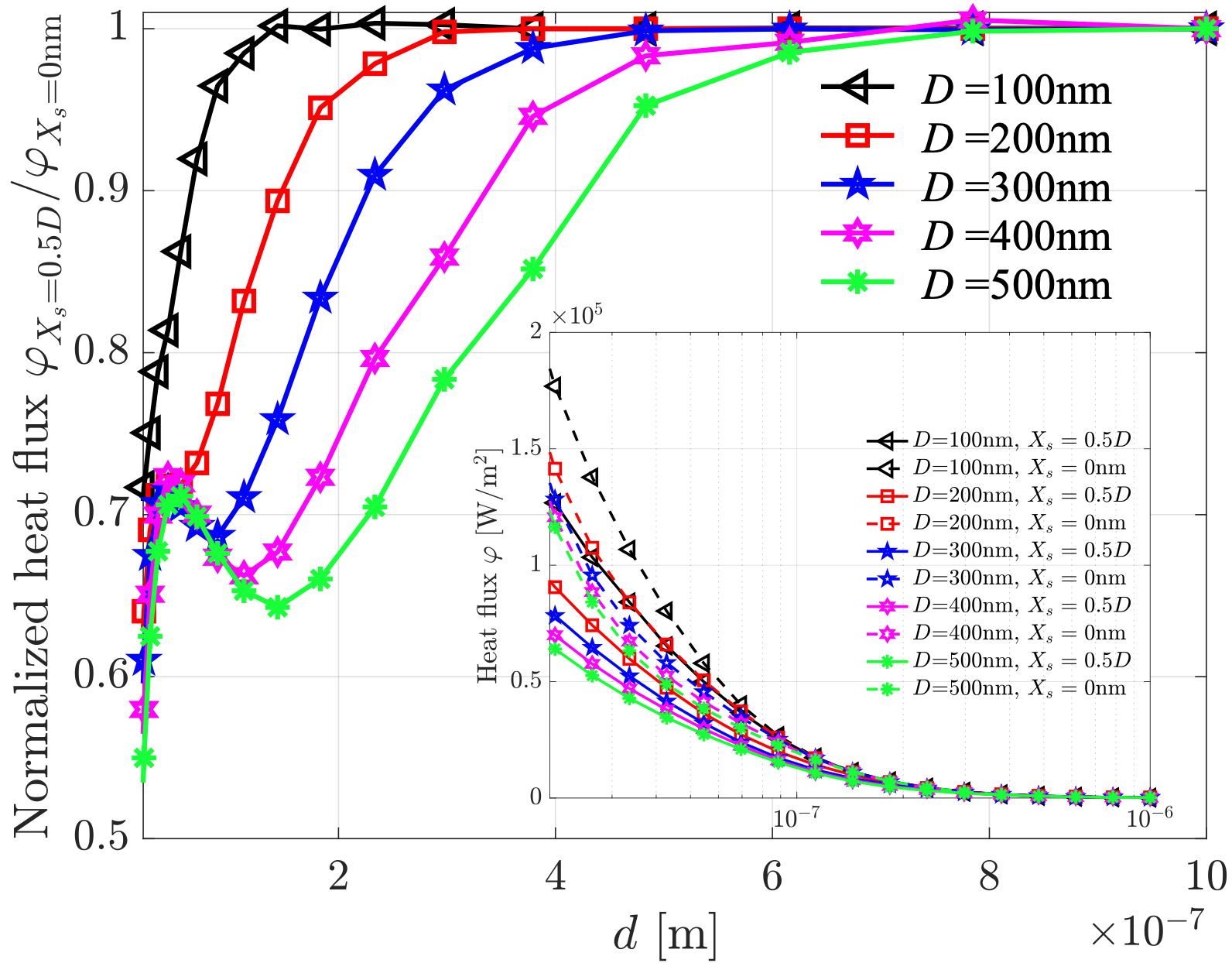}}
     \caption{Dependence of the ratio $\varphi_{X_s=0.5D}^{}/\varphi_{X_s=0\text{nm}}^{}$ (main figure) and the heat flux $\varphi$ (inset) on separation distance $d$ for five different grating periods, $D=100$nm, 200nm, 300nm, 400nm and 500nm. The filling fraction $f$ = 0.5, chemical potential $\mu=0.5$eV.}
   \label{fig:regime_b_five_lines}
\end{figure}

Additionally, for a fixed separation $d$, as the period $D$ decreases, the ratio $\varphi_{X_s=0.5D}^{}/\varphi_{X_s=0\text{nm}}^{}$ also approaches 1 gradually and the scattering details of the graphene grating do not change with the relative lateral shift $X_s$ anymore. When the grating period $D$ is relatively small compared to the separation $d$, the graphene grating behaves like an effective medium, and the relative lateral shift between the two gratings does not change the scattering details and thus does not affect the heat transfer. We can clearly see that the ratio $\varphi_{X_s=0.5D}^{}/\varphi_{X_s=0\text{nm}}^{}$ generally increases to 1 as the geometric factor $d/D$ increases. The lateral shift effect on heat transfer becomes less and less important as the $d/D$ increases. Two distinct regions can be distinguished, where we see a significant lateral shift effect (LSE) and a negligible lateral shift effect on heat transfer, respectively (namely, the LSE region and the non-LSE region). In the LSE region, the lateral shift can even result in over 50\% reduction of the heat flux. For clarity, we take several lines ($D=100$nm, 200nm, 300nm, 400nm and 500nm) from Fig.~\ref{fig:regime_b} and show them in Fig.~\ref{fig:regime_b_five_lines} to follow the details of the the dependence of the ratio $\varphi_{X_s=0.5D}^{}/\varphi_{X_s=0\text{nm}}^{}$ on the separation $d$. The inset in Fig.~\ref{fig:regime_b_five_lines} is for the dependence of the absolute heat flux on the separation $d$.

%

Although the heat flux $\varphi$ decreases monotonically with the separation $d$ for both $X_s=0$nm and $X_s=0.5D$, the ratio $\varphi_{X_s=0.5D}^{}/\varphi_{X_s=0\text{nm}}^{}$ is usually not monotonic with this separation $d$. The dependence of heat flux $\varphi$ on $d$ is not synchronized for the two configurations with lateral shift $X_s=0$nm and 0.5$D$. For the gratings with the same filling fraction $f$, as the separation $d$ increases to 200 nm or more, they always have the same heat flux, even though they have different periods $D$. That is, for two gratings with a large enough separation, it is the filling fraction that plays the determining role for the heat transfer, rather than the grating period.

\begin{figure*} [htbp]
\centerline {\includegraphics[width=0.8\textwidth]{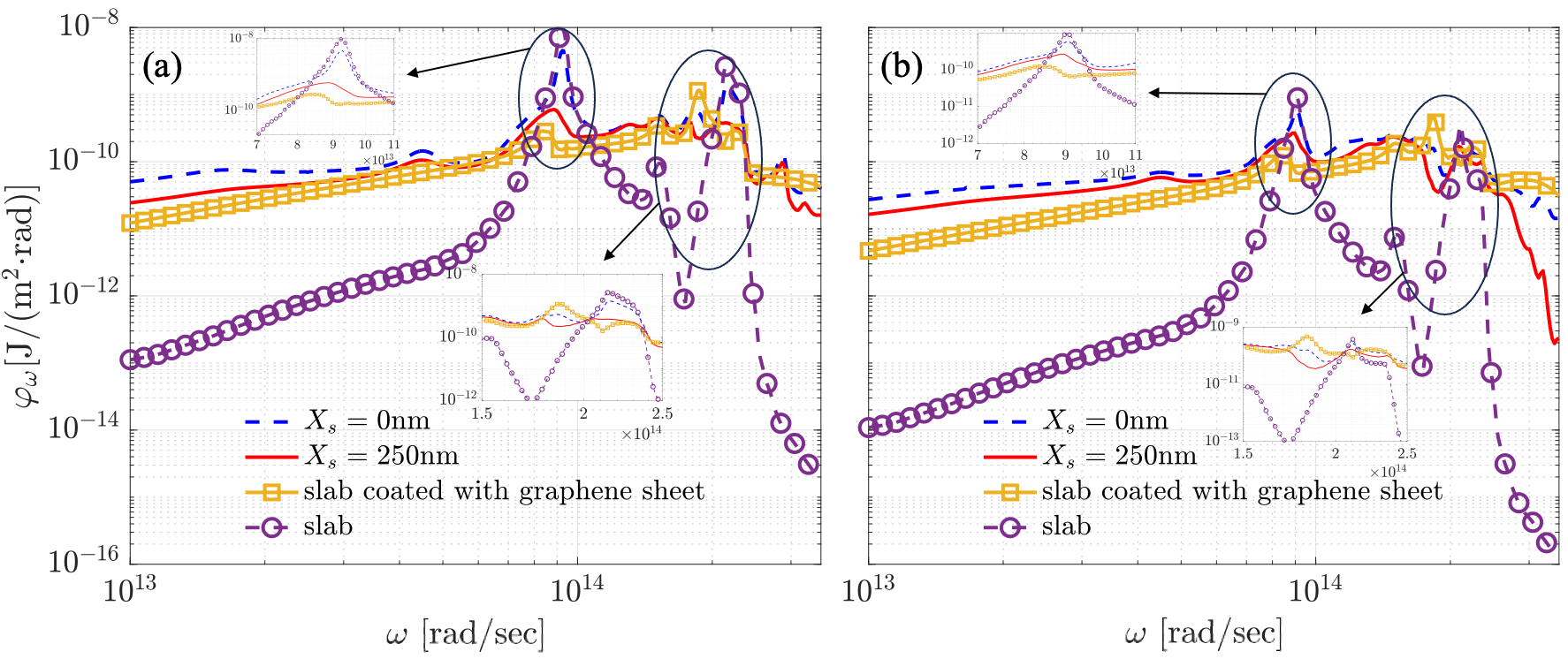}}
     \caption{Spectra of heat flux between two objects with two relative shifts ($X_s=0$nm and $0.5D$) and two different separations: (a) separation $d=20$nm, (b) separation $d=50$nm. The grating period $D=500$nm, filling fraction $f$ = 0.5, chemical potential $\mu=0.5$eV. The heat flux spectrums for the configurations of bare slabs and slabs coated with graphene sheets are also added for reference.}
   \label{fig:regime_b_spectrum}
\end{figure*}

To understand the separation-dependent shift effect on the heat flux, we show its spectrum for two separations ($d=20$nm and 50nm) in Fig. \ref{fig:regime_b_spectrum}. Two relative shifts are considered, $X_s=0$nm and $0.5D$ with $D=500$nm, $f$ = 0.5 and $\mu=0.5$eV. The spectra for the configurations of bare slabs and slabs coated with graphene sheets are also added for reference. The graphene coating (including the graphene grating and the graphene sheet) increases the whole heat flux spectrum apart from the peaks. As compared to the graphene grating coating, the graphene sheet coating can induce a reduction of the low-frequency (at around $9\times10^{13}$rad/s) peak and a redshift of the high-frequency (at around $2\times10^{14}$rad/s) peak. The lateral shift between the two objects reduces the whole spectrum of heat flux from the dashed blue line to the red solid line (as shown in Fig. \ref{fig:regime_b_spectrum}). The difference between the red line and blue line spectra is more pronounced for small $d$. 

\begin{figure*} [htbp]
\centerline {\includegraphics[width=1.0\textwidth]{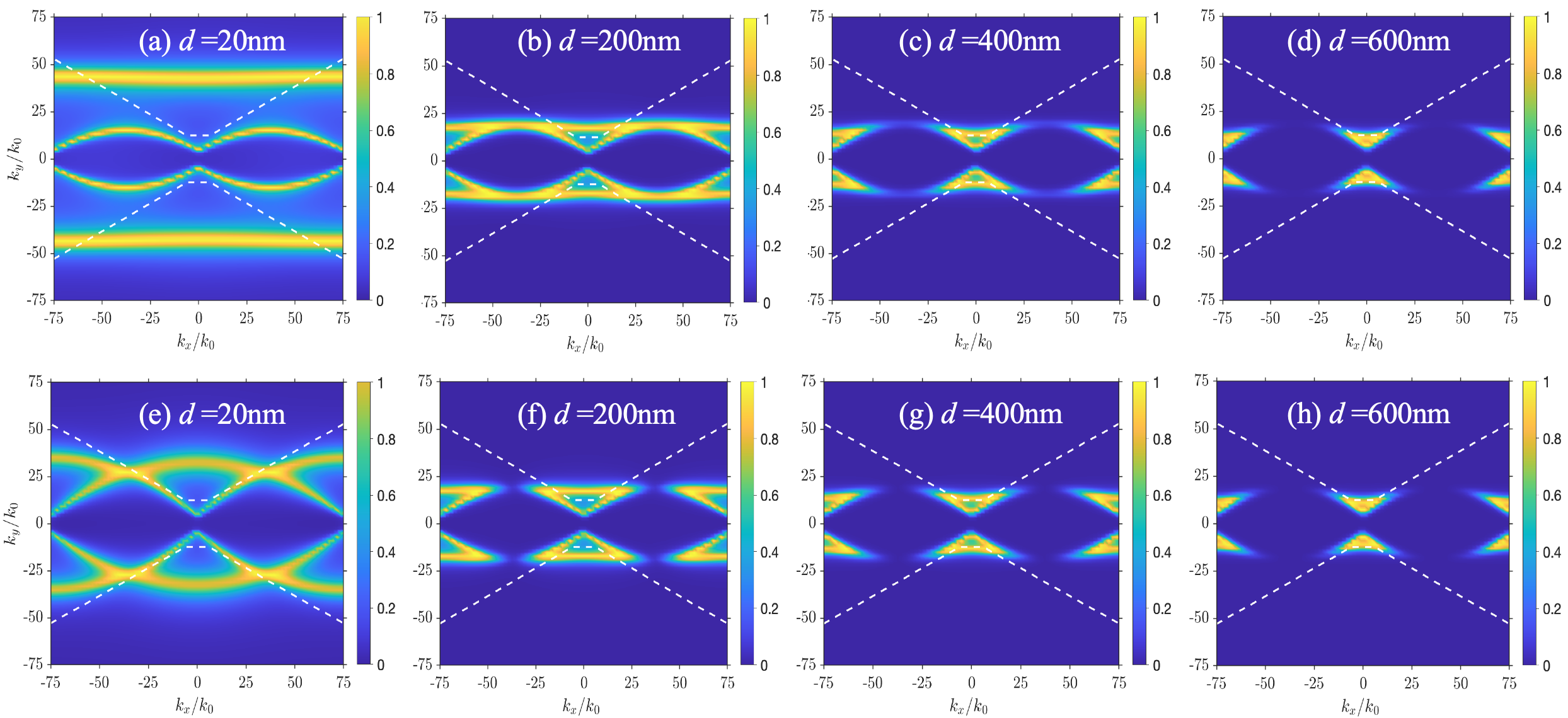}}
     \caption{Energy transmission coefficient for a fused silica substrate coated with a graphene grating: (a) separation $d=20$nm and lateral shift $X_s=0$nm, (b) separation $d=200$nm and lateral shift $X_s=0$nm, (c) separation $d=400$nm and lateral shift $X_s=0$nm, (d) separation $d=600$nm and lateral shift $X_s=0$nm, (e) separation $d=20$nm and lateral shift $X_s=250$nm, (f) separation $d=200$nm and lateral shift $X_s=250$nm, (g) separation $d=400$nm and lateral shift $X_s=250$nm, (h) separation $d=600$nm and lateral shift $X_s=250$nm. The grating period $D=500$nm, chemical potential $\mu=0.5$eV, filling fraction $f$=0.5.}
   \label{fig:regime_b_contour}
\end{figure*}

To further understand the observations concerning the separation-dependent lateral shift effect on the radiative heat transfer in Figs.~\ref{fig:regime_b} and ~\ref{fig:regime_b_five_lines}, we show the energy transmission coefficient Tr($\mathcal{O}$) (see Eq.~(\ref{O_operator}) for the definition) in the ($k_x/k_0$, $k_y/k_0$) plane at the angular frequency $\omega=5\times 10^{13}$rad/s for a slab coated with a graphene grating in Fig.~\ref{fig:regime_b_contour}. Eight configurations are considered where the upper panels (a-b-c-d) are for the configurations without a lateral shift (\textit{i.e.}, $X_s=0$nm), and the lower panels (e-f-g-h) are for configurations with a fixed half-period lateral shift (\textit{i.e.}, $X_s=0.5D=250$nm). The separations $d=20$nm, 200nm, 400nm and 600nm are considered while $D=500$nm, $\mu=0.5$eV, and $f$=0.5. The dotted curves represent the dispersion relations obtained from the poles of the reflection coefficient \cite{Zhou2022langmuir,Zhou2022prm}. 
%
%

When increasing the separation $d$, by comparing the panels (a-b-c-d) to panels (e-f-g-h) correspondingly in Fig.~\ref{fig:regime_b_contour}, we show that the difference between the topologies of the allowed modes caused by the relative lateral shift  becomes less and less important. In addition, we can see in Fig.~\ref{fig:regime_b_contour} a shrinking region for the allowed modes as the separation $d$ increases, regardless of whether there is a lateral shift or not, which results in a decreasing heat flux. Particularly, at separation $d=20$nm, the high-$k$ branch allowed modes blends and hybridizes significantly with the relative low-$k$ branch due to the difference in the near-field caused by the lateral shift. The near-field effect weakens with the mismatch between the two objects. Hence, less energy can be exchanged between the two objects due to the lateral shift at short separation. A general conclusion is that the lateral shift works against the radiative heat transfer for short separations but becomes less important for large separations.

\section{conclusion}

We analyzed theoretically the effect of a lateral shift on near-field radiative heat transfer between finite-thickness planar fused silica slabs coated with graphene gratings, through the Fourier modal method augmented with local basis functions (FMM-LBF). Compared to the effective medium approximation, which treats the grating as an effective whole rather than in detail, this more accurate approach goes far beyond, especially because the former cannot account for the lateral shift. 
In order to take into account the lateral shift in the FMM-LBF and  besides the existing method of including the lateral shift by directly applying the reference translation to the scattering operators \cite{Messina2011PRA}, we developed another method to include the lateral shift effect in the scattering operators themselves.

We show that, due to the lateral shift, the heat flux can exhibit significant oscillations and even a 60\%-70\% reduction compared to the aligned case {(\textit{e.g.},  $f = 0.2$, $\mu = 0.5$eV, $D = 1000$nm, $d = 100$nm)}. 
 Such a lateral shift effect is found to be sensitive to the geometric factor $d/D$. If $d/D<0.5$ (see the Fig.~\ref{fig:regime_b} for the regime map of the lateral shift effect), the ratio of the heat flux with a half period shift to that without shift ($\varphi_{X_s=0.5D}^{}/\varphi_{X_s=0\text{nm}}^{}$) is far from unity. That is, the lateral shift reduces significantly the heat flux. However, we can generally say that when { $d/D>1$ } the ratio $\varphi_{X_s=0.5D}^{}/\varphi_{X_s=0\text{nm}}^{}$ {is close to 1 (see Fig.~\ref{fig:d_D_1} and Fig.~\ref{fig:regime_b})}, where the lateral shift effect on heat transfer becomes less important.
We can clearly distinguish two asymptotic regimes for the radiative heat transfer, \textit{i.e.}, the LSE regime and the non-LSE regime, where we see a significant lateral shift effect and a negligible lateral shift effect on heat transfer, respectively. 
Regardless of the lateral shift, the radiative heat flux shows a non-monotonic dependence on the graphene chemical potential, with the heat flux reaching its maximum value at  $\mu \approx $0.25eV. 
Considering that all the length scales (\textit{e.g.}, the lateral shift and the period) are realistic and experimentally accessible, the lateral shift sensitive and chemical potential dependent radiative heat transfer might have potential for the thermal logic gate. It would also be interesting to study the effect of the grating's lateral shift on Casimir interactions \cite{Casimir_Noto,Hobun}.

\begin{acknowledgments}
This work described was supported by a grant "CAT" (No. A-HKUST604/20) from the ANR/RGC Joint Research Scheme sponsored by the French National Research Agency (ANR) and the Research Grants Council (RGC) of the Hong Kong Special Administrative Region, China.
\end{acknowledgments}

\appendix*
\section{Modified FMM-LBF with a lateral shift}
\label{Mod_LBF}

The main text outlined two distinct approaches  to compute the scattering matrix of our structure. In the first approach, the lateral shift is directly incorporated into the scattering matrix elements (as presented in Eq. \eqref{R2_T2_}). In contrast,  the second approach requires a modification of the FMM-LBF method itself. This appendix is dedicated to a detailed derivation of the latter approach.

In our recent work \cite{Luo2023Casimir_gg}, we provided comprehensive details of the FMM-LBF with no lateral shift ($X_s = 0$). In this context, we initially computed the interface scattering matrix $S_{\rm LBF}$ between the input medium I and medium II (as shown in Fig.~\ref{fig:shift_diagram}). Then, we determined the slab scattering matrix, denoted $S_{\text{slab}}$, between medium II and medium III. By performing the star product ($\star$) operation \cite{Luo2023Casimir_gg}, between $S_{\rm LBF}$ and $S_{\text{slab}}$, we obtained the overall scattering matrix, denoted $S$ and given by:
\begin{equation}
S = S_{\rm LBF} \star S_{\rm slab}=\begin{pmatrix}
\mathcal{R}_{xyz}^{-} & \mathcal{T}_{xyz}^{-}\\
\mathcal{T}_{xyz}^{+} & \mathcal{R}_{xyz}^{+}
\end{pmatrix},
\label{star_product_S}
\end{equation}
where $\mathcal{R}_{xyz}^{+}$ and $\mathcal{R}_{xyz}^{-}$ ($\mathcal{T}_{xyz}^{+}$ and $\mathcal{T}_{xyz}^{-}$) are the reflection operators (transmission operators) in the $(x, y, z)$ Cartesian basis.

In the presence of a lateral shift $X_s$, as illustrated in Fig.~\ref{fig:shift_diagram}, the definition of the matrix $S_{\rm LBF}$ is modified, while the rest of the calculation process remains the same. Let us proceed to compute this new matrix.

Due to periodicity along the $x$ direction, new diffraction channels open up, characterized by the wave vector component in that direction. The $z$-component of the $n^{\rm th}$ diffraction order wave vector depends on the medium. $k_{zn}^{\rm I}$ and $k_{zn}^{\rm II}$ represent the $z$ wave vector of the $n^{\rm th}$ diffraction order for medium I (with $\varepsilon_{\rm I}$ on the incidence side) and medium II (with $\varepsilon_{\rm II}$ on the output side), respectively, and are given by:
\begin{equation}
\left\{
\begin{aligned}
k_{zn}^{\rm I}=\sqrt{k_0^2\varepsilon_{\rm I}- k_{xn}^2- k_y^2}, \\
k_{zn}^{\rm II}=\sqrt{k_0^2\varepsilon_{\rm II}- k_{xn}^2- k_y^2} ,
\end{aligned}
\right.
\label{gamma_I_II}
\end{equation}
with $k_0 = \omega/c$, $k_{xn}=k_x+n \frac{2\pi}{D}$, $k_{x}$ is in the first Brillouin zone $(-\frac{\pi}{D},\frac{\pi}{D})$, and $k_y$ is in $\mathbb{R}$. 

The electric field and the magnetic field in medium I can be expressed as
\begin{equation}
\left\{
\begin{aligned}
\textbf{E}_{\rm I}&=\sum_{n} \left(\textbf{I}_n e^{i{\textbf{k}}_{{\rm i}n}\cdot \textbf{r}}+\textbf{R}_n e^{i{\textbf{k}}_{{\rm r}n}\cdot \textbf{r}}\right),\\
\textbf{H}_{\rm I}&=\frac{1}{k_0 Z_0}\sum_n \left({\textbf{k}}_{{\rm i}n} \times \textbf{I}_n e^{i{\textbf{k}}_{{\rm i}n}\cdot \textbf{r}}+{\textbf{k}}_{{\rm r}n} \times \textbf{R}_n e^{i{\textbf{k}}_{{\rm r}n}\cdot \textbf{r}}\right),
\end{aligned}
\right.
\label{Conical_E_field_I}
\end{equation}
Here, $\textbf{I}_n=(I_{xn},I_{yn},I_{zn})$, $\textbf{R}_n=(R_{xn},R_{yn},R_{zn})$, ${\textbf{k}}_{{\rm i}n}=( k_{xn}, k_y, k_{zn}^{\rm I})$, ${\textbf{k}}_{{\rm r}n}=( k_{xn}, k_y,- k_{zn}^{\rm I})$ and $Z_0=\sqrt{\dfrac{\mu_0}{\varepsilon_0}}$, with $n \in \mathbb{Z}$. Typically, in the numerical implementation, we retain only $2N+1$ Fourier coefficients {\it i.e.} $n \in [-N,N]$; where $N$ is called the truncation order.

In medium II, the electric field and the magnetic field are:
\begin{equation}
\left\{
\begin{aligned}
\textbf{E}_{\rm II}&=\sum_n \left(\textbf{T}_n e^{i{\textbf{k}}_{{\rm t}n}\cdot \textbf{r}}+\textbf{I}_n' e^{i{\textbf{k}}_{{\rm i'}n}\cdot \textbf{r}}\right),\\
\textbf{H}_{\rm II}&=\frac{1}{k_0 Z_0}\sum_n \left({\textbf{k}}_{{\rm t}n} \times \textbf{T}_n e^{i{\textbf{k}}_{{\rm t}n}\cdot \textbf{r}}+{\textbf{k}}_{{\rm i'}n} \times \textbf{I}_n' e^{i{\textbf{k}}_{{\rm i'}n}\cdot \textbf{r}}\right),
\end{aligned}
\right.
\label{Conical_E_field_II}
\end{equation}
where $\textbf{T}_n=(T_{xn},T_{yn},T_{zn})$, $\textbf{I}_n'=(I_{xn}',I_{yn}',I_{zn}')$, ${\textbf{k}}_{{\rm t}n}=( k_{xn}, k_y, k_{zn}^{\rm II})$ and ${\textbf{k}}_{{\rm i'}n}=( k_{xn}, k_y,- k_{zn}^{\rm II})$. 

Furthermore, using ${\rm div}\textbf{E}=\textbf{k} \cdot \textbf{E}=0$, we have the following relations
\begin{equation}
\begin{aligned}
& I_{zn}=-\frac{1}{ k_{zn}^{\rm I}}( k_{xn} I_{xn}+ k_yI_{yn}), \\
&R_{zn}=\frac{1}{ k_{zn}^{\rm I}}( k_{xn} R_{xn}+ k_y R_{yn}),
\\& T_{zn}=-\frac{1}{ k_{zn}^{\rm II}}( k_{xn} T_{xn}+ k_yT_{yn}), \\
&I_{zn}'=\frac{1}{ k_{zn}^{\rm II}}( k_{xn} I_{xn}'+ k_y I_{yn}').
\label{Iz_Rz_Tz_Izprime}
\end{aligned}
\end{equation}

The boundary conditions for the electric field at $z=0$ are
\begin{equation}
\left\{
\begin{aligned}
E_{{\rm I}x}(x,y,0)=E_{{\rm II}x}(x,y,0),\\
E_{{\rm I}y}(x,y,0)=E_{{\rm II}y}(x,y,0).
\end{aligned}
\right.
\label{conical_Exy_boundary}
\end{equation}

By inserting Eqs. (\ref{Conical_E_field_I}) and (\ref{Conical_E_field_II}) into Eq. (\ref{conical_Exy_boundary}),  for arbitrary $n$, we have

\begin{equation}
\left\{
\begin{aligned}
I_{xn}+R_{xn}=I_{xn}'+T_{xn},\\
I_{yn}+R_{yn}=I_{yn}'+T_{yn},
\end{aligned}
\right.
\label{E_n}
\end{equation}
which can be expressed  in compact form
\begin{equation}
I+R=I'+T,
\label{conical_E_boundary}
\end{equation}
where 
\begin{equation}
I=\begin{pmatrix}
I_x\\
I_y
\end{pmatrix},R=\begin{pmatrix}
R_x\\
R_y
\end{pmatrix},I'=\begin{pmatrix}
I_x'\\
I_y'
\end{pmatrix},T=\begin{pmatrix}
T_x\\
T_y
\end{pmatrix}.
\label{Ix_Iy}
\end{equation}

Due to the zero thickness approximation of the graphene grating, the boundary condition for the magnetic fields at the interface between media I and II are
\begin{equation}
H_{{\rm II}x}(x,y,0)-H_{{\rm I}x}(x,y,0)=\sigma(x) E_{{\rm II}y}(x,y,0),
\label{1st_BDC}
\end{equation}
where the function $\sigma(x)$ is periodic and can be expanded into Fourier series as follows
\begin{equation}
\sigma(x) = \begin{cases}
\sigma_g &\text{ if  }  \;\;X_s < x < X_s + a~\textrm{(graphene)} \\
0 &\text{ if }  \;\; X_s + a < x < X_s + D~\textrm{(slit)} 
\end{cases} = \sum_{n} \sigma_{n} e^{i\frac{2\pi}{D}nx}.
\end{equation}

By inserting Eqs. (\ref{Conical_E_field_I})and (\ref{Conical_E_field_II}) into Eq. (\ref{1st_BDC}), and using the Laurent factorization rule, the following relation is obtained
\begin{equation}
\begin{aligned}
& \left\{( k_y T_{zn}- k_{zn}^{\rm II} T_{yn})+( k_y I_{zn}'+ k_{zn}^{\rm II} I_{yn}')\right.\\
&\left.-( k_y I_{zn}- k_{zn}^{\rm I} I_{yn})-( k_y R_{zn}+ k_{zn}^{\rm I} R_{yn})\right\}\\
&=k_0 Z_0 \sum_{n'}\left\{\sigma_{n'-n} (T_{yn'}+I_{yn'}')\right\}.
\end{aligned}
\label{1_Hx_boundary_n}
\end{equation}

That can be expressed in a compact matrix form as

\begin{equation}
\begin{aligned}
&( k_y T_{z}-\gamma_{\rm II} T_{y})+( k_y I_{z}'+\gamma_{\rm II} I_{y}')-( k_y I_{z}-\gamma_{\rm I} I_{y})\\
&-( k_y R_{z}+\gamma_{\rm I} R_{y})=k_0 Z_0 [[\sigma]] (T_{y}+I_{y}'),
\end{aligned}
\label{1_Hx_boundary_matrix}
\end{equation}
where $\gamma_{\rm II}=$diag($k_{zn}^{\rm II}$), $\gamma_{\rm I}=$diag($k_{zn}^{\rm I}$) and $[[\sigma]]$ is the Toeplitz matrix whose ($n$', $n$) element is $ \sigma_{n'-n} $, more precisely: 
\begin{equation}
\begin{aligned}
[[\sigma]]=
\begin{pmatrix}
\sigma_0 & \sigma_{-1} & & &  \sigma_{-2N}\\
\sigma_{1} & \sigma_0 & \sigma_{-1} & &  &\\
& \ddots & \ddots & \ddots &\\ & & \ddots & \ddots & \sigma_{-1} &\\ \sigma_{2N}^{}& & &  \sigma_{1}& \sigma_0& 
\end{pmatrix},
\label{topleitz_matrix}
\end{aligned}
\end{equation}

\noindent where $\sigma_p$=$\frac{i\sigma_g }{2\pi p} e^{-\frac{i2 \pi p X_s} {D}} \left(e^{-\frac{i2 \pi p a}{D} }- 1\right)=e^{-\frac{i2 \pi p X_s} {D}} \overline{\sigma}_p$ when $p$ is a non-zero integer, $\sigma_0=\sigma_g a/D=\overline{\sigma}_0$, $\overline{\sigma}_p$ and $\overline{\sigma}_0$ correspond to the case $X_s=0$ \cite{Luo2023Casimir_gg,Hwang2020}.

In addition, when considering a lateral translation $X_s$, the electric field $E_x$ on the graphene grating surface ($z=0$) can be expressed in terms of local basis functions [$g'_m(x)$ and $s'_m(x)$] as follows 
\begin{equation}
E_{x}(x,y)=e^{i k_yy}\left\{
\begin{aligned}
&\sum_{m=1}^{N_g}p_m g'_m(x) \;\;\;\text{if}\;\;  X_s < x < X_s + a \\
&\sum_{m=0}^{N_s-1}q_m s'_m(x) \;\;\;\text{if}\;\; X_s + a < x < X_s + D \\
\end{aligned}
\right.,
\label{E_x_FMM_LBF_shift}
\end{equation}
where 
\begin{equation}
\;\;\;\;\;\;\;\;\;\;\;\;\;\;\;\;\;\;\;\;\;\;\;\left\{
\begin{aligned}
&g_m(x) = \sin(m \pi x/a) \\\\
&s_m(x) = \dfrac{\cos(m\pi(x-a)/c')}{\sqrt{(c'/2)^2-(x-x_c)^2}} 
\end{aligned}
\right. ,
\label{LBFs}
\end{equation}
with $c'=D-a$, $x_c = (a+D)/2$, $N_g ={ \rm round}\left[ \dfrac{N\times a}{D}\right]$, $N_s = N-N_g$, and round$(x)$ gives the nearest integer number, then
\begin{equation}
\left\{
\begin{aligned}
&g'_m(x) = g_m(x-X_s)\;\; {\rm if}~~X_s < x < X_s + a \\\\
&s'_m(x) = s_m(x-X_s)\;\; {\rm if}~~X_s + a < x < X_s + D  
\end{aligned}
\right. .
\label{LBFs_shift}
\end{equation}

Projecting the boundary condition for the {$y$-component of the} magnetic field on the $(e^{-i k_{xn} x})$ basis we obtain:
%
\begin{equation}
\begin{aligned}
&( k_{zn}^{\rm II} T_{xn}- k_{xn} T_{zn}- k_{zn}^{\rm II} I_{xn}'- k_{xn} I_{zn}')\\&-( k_{zn}^{\rm I} I_{xn}- k_{xn} I_{zn}- k_{zn}^{\rm I} R_{xn}- k_{xn} R_{zn})
\\&=  -\sigma_g k_0 Z_0 \sum_{m=1}^{N_g} <e^{-i k_{xn} x},p_m g'_m(x) >,
\label{conical_Hy_boundary_n_single_1}
\end{aligned}
\end{equation}
%
where the scalar product is given by $<f,g> = \frac{1}{D}\int_{X_s}^{X_s+D}f(x)g(x){\rm d}x$. Then Eq.~(\ref{conical_Hy_boundary_n_single_1}) yields : 
\begin{equation}
\begin{aligned}
&( k_{zn}^{\rm II} T_{xn}- k_{xn} T_{zn}- k_{zn}^{\rm II} I_{xn}'- k_{xn} I_{zn}')\\&-( k_{zn}^{\rm I} I_{xn}- k_{xn} I_{zn}- k_{zn}^{\rm I} R_{xn}- k_{xn} R_{zn})
\\&=-\frac{1}{D}\int_{X_s}^{X_s+a}  \sigma_g k_0 Z_0 \sum_{m=1}^{N_g}p_m g_m(x-X_s) e^{-i k_{xn} x} {\rm d}x.
\label{conical_Hy_boundary_n_single_2}
\end{aligned}
\end{equation}
%
Using the change of variable $x' = x - X_s$, Eq.~(\ref{conical_Hy_boundary_n_single_2}) becomes
\begin{equation}
\begin{aligned}
&( k_{zn}^{\rm II} T_{xn}- k_{xn} T_{zn}- k_{zn}^{\rm II} I_{xn}'- k_{xn} I_{zn}')\\&-( k_{zn}^{\rm I} I_{xn}- k_{xn} I_{zn}- k_{zn}^{\rm I} R_{xn}- k_{xn} R_{zn})
\\&=-\frac{1}{D}\int_{0}^{a}  \sigma_g k_0 Z_0 \sum_{m=1}^{N_g}p_m g_m(x') e^{-i k_{xn} x'-i k_{xn} X_s} {\rm d}x'.
\label{conical_Hy_boundary_n_single_2_value_change}
\end{aligned}
\end{equation}
%
{By taking out the term $e^{-i k_{xn} X_s}$ in the integral} and exchanging the order of summation and integration, the RHS of Eq. (\ref{conical_Hy_boundary_n_single_2_value_change}) becomes
\begin{equation}
\begin{aligned}
&-\frac{1}{D}\int_{0}^{a}  \sigma_g k_0 Z_0 \sum_{m=1}^{N_g}p_m g_m(x') e^{-i k_{xn} x'-i k_{xn} X_s} {\rm d}x'\\&=-\sigma_g k_0 Z_0 e^{-i k_{xn} X_s} \sum_{m=1}^{N_g}p_m G_{nm},
\label{conical_Hx_boundary_n_single_RHS}
\end{aligned}
\end{equation}
where $G_{nm}= \frac{-ia}{2D} e^{-i k_{xn} a/2}\left[ \right.  e^{im\pi/2} sinc (\alpha_{nm}^- a/2)-$ $ e^{-im\pi/2}$   $sinc(\alpha_{nm}^+ a/2) \left.\right]$, $\alpha_{nm}^\pm = m\pi/a \pm  k_{xn}$.\\

Eq.~(\ref{conical_Hy_boundary_n_single_2}) can now be recast in a more compact form :
\begin{equation}
\begin{aligned}
&\gamma_{\rm II} T_{x}-\alpha T_{z}-\gamma_{\rm II} I_{x}'-\alpha I_{z}'-\gamma_{\rm I} I_{x}+\alpha I_{z}\\&+\gamma_{\rm I} R_{x}+\alpha R_{z}=-\sigma_g k_0 Z_0 \Delta \mathbb{G} p,
\label{conical_Hy_boundary_matrix1}
\end{aligned}
\end{equation}
where $\Delta=\text{diag}(e^{-i k_{xn} X_s})$, $\mathbb{G}=\{G_{nm}\}$ is a matrix with size [$(2N+1) \times N_g$] and $p$ is the column vector formed by the $N_g$ coefficients $p_m$ with $p=(p_1^{},p_2^{},p_3^{} \cdots p_{N_g}^{})^{\rm T}$. We can write the above equation as follows:
\begin{equation}
\begin{aligned}
&\gamma_{\rm II} T_{x}-\alpha T_{z}-\gamma_{\rm II} I_{x}'-\alpha I_{z}'-\gamma_{1} I_{x}+\alpha I_{z}+\gamma_{1} R_{x}+\alpha R_{z}\\&=-\sigma_g k_0 Z_0 \left[ \Delta\mathbb{G} \; \textbf{0} \right] \begin{pmatrix} p\\ q\end{pmatrix},
\label{conical_Hy_boundary_matrix1_a_b}
\end{aligned}
\end{equation}
where $\left[ \Delta \mathbb{G} \; \textbf{0} \right]$ is the  horizontale concatenation of matrices $\Delta \mathbb{G}$ and the matrix \textbf{0} denoting the zero matrix of size [$(2N+1) \times N_s$], $q$ is the column vector formed by the $N_s$ coefficients $q_m$ with $q=(q_0^{},q_1^{},q_2^{} \cdots q_{N_s-1}^{})^{\rm T}$.

To obtain $(p,q)^{\rm T}$, we use the boundary condition on the $x$-component of the electric field; and following the same procedure, as that for the $y$-component of the magnetic field, we obtain
\begin{equation}
\begin{aligned}
\begin{pmatrix} p\\ q\end{pmatrix}=\left[\Delta \mathbb{G} \;\Delta \mathbb{S} \right]^{-1} (T_{x}+ I_{x}'),
\label{a_b_matrix}
\end{aligned}
\end{equation}
where $\left[ \Delta \mathbb{G} \; \Delta \mathbb{S} \right]$ is the horizontale concatenation of matrices $\Delta \mathbb{G}$ and  $\Delta \mathbb{S}$, $S_{nm}=\frac{\pi}{2D}e^{-i k_{xn} x_c}\left[ \right. e^{im\pi/2} J_0(\beta_{nm}^- c'/2) +  e^{-im\pi/2} J_0(\beta_{nm}^+ c'/2) \left. \right]$, $\beta_{nm}^\pm = m\pi/c' \pm k_{xn}$, $J_0(x)$ is the zero-order Bessel function of the first kind.

Finally the modified interface scattering matrix $S_{\rm LBF}$ is given by
\begin{equation}
\begin{pmatrix}
R\\
T
\end{pmatrix}=S_{\rm LBF} \begin{pmatrix}
I\\
I'
\end{pmatrix},
\label{Conical_Total_fields}
\end{equation}
where
\begin{equation}
S_{\rm LBF}=\begin{pmatrix}
\mathbbm{1} & -\mathbbm{1}\\
B & A+\Lambda'
\end{pmatrix}^{-1} \begin{pmatrix}
-\mathbbm{1} & \mathbbm{1}\\
B & A-\Lambda'
\end{pmatrix},
\label{S_LBF_conical}
\end{equation}
$\mathbbm{1}$ being the identity matrix of size $[2(2N+1) \times 2(2N+1)]$, $\Lambda'={ \text{diag} (\sigma k_0 Z_0 \left[ \mathbb{G}' \; 0 \right] \left[ \mathbb{G}' \;\mathbb{S}' \right]^{-1},[[\sigma]] k_0 Z_0)}$, $A$ and $B$ are defined as:
\begin{equation}
\begin{aligned}
&A=\begin{pmatrix}
\gamma_{\rm II}+\alpha^2 {\gamma_{\rm II}^{-1}} & {\alpha  k_y}{\gamma_{\rm II}^{-1}}\\
\\
{\alpha  k_y}{\gamma_{\rm II}^{-1}} & \gamma_{\rm II} +{ k_y^2}{\gamma_{\rm II}^{-1}}
\end{pmatrix}, 
\\&B=\begin{pmatrix}
\gamma_{\rm I}+{\alpha^2}{\gamma_{\rm I}^{-1}} & {\alpha  k_y}{\gamma_{\rm I}^{-1}}\\
\\
{\alpha  k_y}{\gamma_{\rm I}^{-1}} & \gamma_{\rm I} +{ k_y^2}{\gamma_{\rm I}^{-1}}
\end{pmatrix}.
\label{A_B_matrix}
\end{aligned}
\end{equation}

As mentioned in the beginning of this appendix, the remaining calculations necessary to obtain the scattering matrix of the global structure [{\it i.e.} the slab scattering matrix $S_{\rm slab}$ and the transformation matrices from the $(x, y, z)$ Cartesian basis to the (TE, TM) basis] proceed exactly as published in our previous work \cite{Luo2023Casimir_gg}. 


\section*{}

\bibliography{HT}

\providecommand{\noopsort}[1]{}\providecommand{\singleletter}[1]{#1}%
\begin{thebibliography}{65}%
\makeatletter
\providecommand \@ifxundefined [1]{%
 \@ifx{#1\undefined}
}%
\providecommand \@ifnum [1]{%
 \ifnum #1\expandafter \@firstoftwo
 \else \expandafter \@secondoftwo
 \fi
}%
\providecommand \@ifx [1]{%
 \ifx #1\expandafter \@firstoftwo
 \else \expandafter \@secondoftwo
 \fi
}%
\providecommand \natexlab [1]{#1}%
\providecommand \enquote  [1]{``#1''}%
\providecommand \bibnamefont  [1]{#1}%
\providecommand \bibfnamefont [1]{#1}%
\providecommand \citenamefont [1]{#1}%
\providecommand \href@noop [0]{\@secondoftwo}%
\providecommand \href [0]{\begingroup \@sanitize@url \@href}%
\providecommand \@href[1]{\@@startlink{#1}\@@href}%
\providecommand \@@href[1]{\endgroup#1\@@endlink}%
\providecommand \@sanitize@url [0]{\catcode `\\12\catcode `\$12\catcode
  `\&12\catcode `\#12\catcode `\^12\catcode `\_12\catcode `\%12\relax}%
\providecommand \@@startlink[1]{}%
\providecommand \@@endlink[0]{}%
\providecommand \url  [0]{\begingroup\@sanitize@url \@url }%
\providecommand \@url [1]{\endgroup\@href {#1}{\urlprefix }}%
\providecommand \urlprefix  [0]{URL }%
\providecommand \Eprint [0]{\href }%
\providecommand \doibase [0]{https://doi.org/}%
\providecommand \selectlanguage [0]{\@gobble}%
\providecommand \bibinfo  [0]{\@secondoftwo}%
\providecommand \bibfield  [0]{\@secondoftwo}%
\providecommand \translation [1]{[#1]}%
\providecommand \BibitemOpen [0]{}%
\providecommand \bibitemStop [0]{}%
\providecommand \bibitemNoStop [0]{.\EOS\space}%
\providecommand \EOS [0]{\spacefactor3000\relax}%
\providecommand \BibitemShut  [1]{\csname bibitem#1\endcsname}%
\let\auto@bib@innerbib\@empty
\bibitem [{\citenamefont {Rytov}\ \emph {et~al.}(1989)\citenamefont {Rytov},
  \citenamefont {Kravtsov},\ and\ \citenamefont {Tatarskii}}]{Rytov1989}%
  \BibitemOpen
  \bibfield  {author} {\bibinfo {author} {\bibfnamefont {S.~M.}\ \bibnamefont
  {Rytov}}, \bibinfo {author} {\bibfnamefont {Y.~A.}\ \bibnamefont
  {Kravtsov}},\ and\ \bibinfo {author} {\bibfnamefont {V.~I.}\ \bibnamefont
  {Tatarskii}},\ }\href@noop {} {\emph {\bibinfo {title} {Priniciples of
  statistical radiophysics}}},\ Vol.~\bibinfo {volume} {3}\ (\bibinfo
  {publisher} {Springer-Verlag},\ \bibinfo {year} {1989})\BibitemShut {NoStop}%
\bibitem [{\citenamefont {Polder}\ and\ \citenamefont
  {Van~Hove}(1971)}]{Polder1971}%
  \BibitemOpen
  \bibfield  {author} {\bibinfo {author} {\bibfnamefont {D.}~\bibnamefont
  {Polder}}\ and\ \bibinfo {author} {\bibfnamefont {M.}~\bibnamefont
  {Van~Hove}},\ }\bibfield  {title} {\bibinfo {title} {Theory of radiative heat
  transfer between closely spaced bodies},\ }\href
  {https://doi.org/10.1103/PhysRevB.4.3303} {\bibfield  {journal} {\bibinfo
  {journal} {Phys. Rev. B}\ }\textbf {\bibinfo {volume} {4}},\ \bibinfo {pages}
  {3303} (\bibinfo {year} {1971})}\BibitemShut {NoStop}%
\bibitem [{\citenamefont {Chapuis}\ \emph
  {et~al.}(2008{\natexlab{a}})\citenamefont {Chapuis}, \citenamefont {Laroche},
  \citenamefont {Volz},\ and\ \citenamefont {Greffet}}]{Chapuis2008plate}%
  \BibitemOpen
  \bibfield  {author} {\bibinfo {author} {\bibfnamefont {P.~O.}\ \bibnamefont
  {Chapuis}}, \bibinfo {author} {\bibfnamefont {M.}~\bibnamefont {Laroche}},
  \bibinfo {author} {\bibfnamefont {S.}~\bibnamefont {Volz}},\ and\ \bibinfo
  {author} {\bibfnamefont {J.-J.}\ \bibnamefont {Greffet}},\ }\bibfield
  {title} {\bibinfo {title} {Near-field induction heating of metallic
  nanoparticles due to infrared magnetic dipole contribution},\ }\href
  {https://doi.org/10.1103/PhysRevB.77.125402} {\bibfield  {journal} {\bibinfo
  {journal} {Phys. Rev. B}\ }\textbf {\bibinfo {volume} {77}},\ \bibinfo
  {pages} {125402} (\bibinfo {year} {2008}{\natexlab{a}})}\BibitemShut
  {NoStop}%
\bibitem [{\citenamefont {Narayanaswamy}\ and\ \citenamefont
  {Chen}(2008)}]{Narayanaswamy2008}%
  \BibitemOpen
  \bibfield  {author} {\bibinfo {author} {\bibfnamefont {A.}~\bibnamefont
  {Narayanaswamy}}\ and\ \bibinfo {author} {\bibfnamefont {G.}~\bibnamefont
  {Chen}},\ }\bibfield  {title} {\bibinfo {title} {Thermal near-field radiative
  transfer between two spheres},\ }\href
  {https://link.aps.org/doi/10.1103/PhysRevB.77.075125} {\bibfield  {journal}
  {\bibinfo  {journal} {Phys. Rev. B}\ }\textbf {\bibinfo {volume} {77}},\
  \bibinfo {pages} {075125} (\bibinfo {year} {2008})}\BibitemShut {NoStop}%
\bibitem [{\citenamefont {Carminati}\ and\ \citenamefont
  {Greffet}(1999)}]{Carminati1999}%
  \BibitemOpen
  \bibfield  {author} {\bibinfo {author} {\bibfnamefont {R.}~\bibnamefont
  {Carminati}}\ and\ \bibinfo {author} {\bibfnamefont {J.-J.}\ \bibnamefont
  {Greffet}},\ }\bibfield  {title} {\bibinfo {title} {Near-field effects in
  spatial coherence of thermal sources},\ }\href
  {https://doi.org/10.1103/PhysRevLett.82.1660} {\bibfield  {journal} {\bibinfo
   {journal} {Phys. Rev. Lett.}\ }\textbf {\bibinfo {volume} {82}},\ \bibinfo
  {pages} {1660} (\bibinfo {year} {1999})}\BibitemShut {NoStop}%
\bibitem [{\citenamefont {Loomis}\ and\ \citenamefont
  {Maris}(1994)}]{Loomis1994}%
  \BibitemOpen
  \bibfield  {author} {\bibinfo {author} {\bibfnamefont {J.~J.}\ \bibnamefont
  {Loomis}}\ and\ \bibinfo {author} {\bibfnamefont {H.~J.}\ \bibnamefont
  {Maris}},\ }\bibfield  {title} {\bibinfo {title} {Theory of heat transfer by
  evanescent electromagnetic waves},\ }\href
  {https://doi.org/10.1103/PhysRevB.50.18517} {\bibfield  {journal} {\bibinfo
  {journal} {Phys. Rev. B}\ }\textbf {\bibinfo {volume} {50}},\ \bibinfo
  {pages} {18517} (\bibinfo {year} {1994})}\BibitemShut {NoStop}%
\bibitem [{\citenamefont {Shchegrov}\ \emph {et~al.}(2000)\citenamefont
  {Shchegrov}, \citenamefont {Joulain}, \citenamefont {Carminati},\ and\
  \citenamefont {Greffet}}]{Shchegrov2000}%
  \BibitemOpen
  \bibfield  {author} {\bibinfo {author} {\bibfnamefont {A.~V.}\ \bibnamefont
  {Shchegrov}}, \bibinfo {author} {\bibfnamefont {K.}~\bibnamefont {Joulain}},
  \bibinfo {author} {\bibfnamefont {R.}~\bibnamefont {Carminati}},\ and\
  \bibinfo {author} {\bibfnamefont {J.-J.}\ \bibnamefont {Greffet}},\
  }\bibfield  {title} {\bibinfo {title} {Near-field spectral effects due to
  electromagnetic surface excitations},\ }\href
  {https://doi.org/10.1103/PhysRevLett.85.1548} {\bibfield  {journal} {\bibinfo
   {journal} {Phys. Rev. Lett.}\ }\textbf {\bibinfo {volume} {85}},\ \bibinfo
  {pages} {1548} (\bibinfo {year} {2000})}\BibitemShut {NoStop}%
\bibitem [{\citenamefont {Volokitin}\ and\ \citenamefont
  {Persson}(2001)}]{Volokitin2001}%
  \BibitemOpen
  \bibfield  {author} {\bibinfo {author} {\bibfnamefont {A.~I.}\ \bibnamefont
  {Volokitin}}\ and\ \bibinfo {author} {\bibfnamefont {B.~N.~J.}\ \bibnamefont
  {Persson}},\ }\bibfield  {title} {\bibinfo {title} {Radiative heat transfer
  between nanostructures},\ }\href {https://doi.org/10.1103/PhysRevB.63.205404}
  {\bibfield  {journal} {\bibinfo  {journal} {Phys. Rev. B}\ }\textbf {\bibinfo
  {volume} {63}},\ \bibinfo {pages} {205404} (\bibinfo {year}
  {2001})}\BibitemShut {NoStop}%
\bibitem [{\citenamefont {Chapuis}\ \emph
  {et~al.}(2008{\natexlab{b}})\citenamefont {Chapuis}, \citenamefont {Laroche},
  \citenamefont {Volz},\ and\ \citenamefont {Greffet}}]{Chapuis2008}%
  \BibitemOpen
  \bibfield  {author} {\bibinfo {author} {\bibfnamefont {P.~O.}\ \bibnamefont
  {Chapuis}}, \bibinfo {author} {\bibfnamefont {M.}~\bibnamefont {Laroche}},
  \bibinfo {author} {\bibfnamefont {S.}~\bibnamefont {Volz}},\ and\ \bibinfo
  {author} {\bibfnamefont {J.-J.}\ \bibnamefont {Greffet}},\ }\bibfield
  {title} {\bibinfo {title} {Radiative heat transfer between metallic
  nanoparticles},\ }\href {https://doi.org/10.1063/1.2931062} {\bibfield
  {journal} {\bibinfo  {journal} {Appl. Phys. Lett.}\ }\textbf {\bibinfo
  {volume} {92}},\ \bibinfo {pages} {3303} (\bibinfo {year}
  {2008}{\natexlab{b}})}\BibitemShut {NoStop}%
\bibitem [{\citenamefont {Manjavacas}\ and\ \citenamefont {Garc\'{\i}a~de
  Abajo}(2012)}]{Manjavacas2012}%
  \BibitemOpen
  \bibfield  {author} {\bibinfo {author} {\bibfnamefont {A.}~\bibnamefont
  {Manjavacas}}\ and\ \bibinfo {author} {\bibfnamefont {F.~J.}\ \bibnamefont
  {Garc\'{\i}a~de Abajo}},\ }\bibfield  {title} {\bibinfo {title} {Radiative
  heat transfer between neighboring particles},\ }\href
  {https://doi.org/10.1103/PhysRevB.86.075466} {\bibfield  {journal} {\bibinfo
  {journal} {Phys. Rev. B}\ }\textbf {\bibinfo {volume} {86}},\ \bibinfo
  {pages} {075466} (\bibinfo {year} {2012})}\BibitemShut {NoStop}%
\bibitem [{\citenamefont {Nikbakht}(2018)}]{Nikbakht2018}%
  \BibitemOpen
  \bibfield  {author} {\bibinfo {author} {\bibfnamefont {M.}~\bibnamefont
  {Nikbakht}},\ }\bibfield  {title} {\bibinfo {title} {Radiative heat transfer
  between core-shell nanoparticles},\ }\href
  {https://doi.org/https://doi.org/10.1016/j.jqsrt.2018.10.005} {\bibfield
  {journal} {\bibinfo  {journal} {J. Quant. Spectrosc. Radiat. Transf.}\
  }\textbf {\bibinfo {volume} {221}},\ \bibinfo {pages} {164} (\bibinfo {year}
  {2018})}\BibitemShut {NoStop}%
\bibitem [{\citenamefont {Messina}\ \emph {et~al.}(2018)\citenamefont
  {Messina}, \citenamefont {Biehs},\ and\ \citenamefont
  {Ben-Abdallah}}]{Messina2018}%
  \BibitemOpen
  \bibfield  {author} {\bibinfo {author} {\bibfnamefont {R.}~\bibnamefont
  {Messina}}, \bibinfo {author} {\bibfnamefont {S.-A.}\ \bibnamefont {Biehs}},\
  and\ \bibinfo {author} {\bibfnamefont {P.}~\bibnamefont {Ben-Abdallah}},\
  }\bibfield  {title} {\bibinfo {title} {Surface-mode-assisted amplification of
  radiative heat transfer between nanoparticles},\ }\href
  {https://doi.org/10.1103/PhysRevB.97.165437} {\bibfield  {journal} {\bibinfo
  {journal} {Phys. Rev. B}\ }\textbf {\bibinfo {volume} {97}},\ \bibinfo
  {pages} {165437} (\bibinfo {year} {2018})}\BibitemShut {NoStop}%
\bibitem [{\citenamefont {Dong}\ \emph {et~al.}(2018)\citenamefont {Dong},
  \citenamefont {Zhao},\ and\ \citenamefont {Liu}}]{DongPrb2018}%
  \BibitemOpen
  \bibfield  {author} {\bibinfo {author} {\bibfnamefont {J.}~\bibnamefont
  {Dong}}, \bibinfo {author} {\bibfnamefont {J.~M.}\ \bibnamefont {Zhao}},\
  and\ \bibinfo {author} {\bibfnamefont {L.~H.}\ \bibnamefont {Liu}},\
  }\bibfield  {title} {\bibinfo {title} {Long-distance near-field energy
  transport via propagating surface waves},\ }\href
  {https://doi.org/10.1103/PhysRevB.97.075422} {\bibfield  {journal} {\bibinfo
  {journal} {Phys. Rev. B}\ }\textbf {\bibinfo {volume} {97}},\ \bibinfo
  {pages} {075422} (\bibinfo {year} {2018})}\BibitemShut {NoStop}%
\bibitem [{\citenamefont {Zhang}\ \emph {et~al.}(2019)\citenamefont {Zhang},
  \citenamefont {Yi}, \citenamefont {Tan},\ and\ \citenamefont
  {Antezza}}]{Zhang2019T}%
  \BibitemOpen
  \bibfield  {author} {\bibinfo {author} {\bibfnamefont {Y.}~\bibnamefont
  {Zhang}}, \bibinfo {author} {\bibfnamefont {H.~L.}\ \bibnamefont {Yi}},
  \bibinfo {author} {\bibfnamefont {H.~P.}\ \bibnamefont {Tan}},\ and\ \bibinfo
  {author} {\bibfnamefont {M.}~\bibnamefont {Antezza}},\ }\bibfield  {title}
  {\bibinfo {title} {Giant resonant radiative heat transfer between
  nanoparticles},\ }\href {https://doi.org/10.1103/PhysRevB.100.134305}
  {\bibfield  {journal} {\bibinfo  {journal} {Phys. Rev. B}\ }\textbf {\bibinfo
  {volume} {100}},\ \bibinfo {pages} {134305} (\bibinfo {year}
  {2019})}\BibitemShut {NoStop}%
\bibitem [{\citenamefont {Luo}\ \emph {et~al.}(2020{\natexlab{a}})\citenamefont
  {Luo}, \citenamefont {Zhao}, \citenamefont {Liu},\ and\ \citenamefont
  {Antezza}}]{Luo2020prb_ensemble}%
  \BibitemOpen
  \bibfield  {author} {\bibinfo {author} {\bibfnamefont {M.~G.}\ \bibnamefont
  {Luo}}, \bibinfo {author} {\bibfnamefont {J.~M.}\ \bibnamefont {Zhao}},
  \bibinfo {author} {\bibfnamefont {L.~H.}\ \bibnamefont {Liu}},\ and\ \bibinfo
  {author} {\bibfnamefont {M.}~\bibnamefont {Antezza}},\ }\bibfield  {title}
  {\bibinfo {title} {Radiative heat transfer and radiative thermal energy for
  two-dimensional nanoparticle ensembles},\ }\href
  {https://doi.org/10.1103/PhysRevB.102.024203} {\bibfield  {journal} {\bibinfo
   {journal} {Phys. Rev. B}\ }\textbf {\bibinfo {volume} {102}},\ \bibinfo
  {pages} {024203} (\bibinfo {year} {2020}{\natexlab{a}})}\BibitemShut
  {NoStop}%
\bibitem [{\citenamefont {Luo}\ \emph {et~al.}(2023{\natexlab{a}})\citenamefont
  {Luo}, \citenamefont {Zhao},\ and\ \citenamefont
  {Liu}}]{Luo2023IJHMT_diffusion}%
  \BibitemOpen
  \bibfield  {author} {\bibinfo {author} {\bibfnamefont {M.~G.}\ \bibnamefont
  {Luo}}, \bibinfo {author} {\bibfnamefont {J.~M.}\ \bibnamefont {Zhao}},\ and\
  \bibinfo {author} {\bibfnamefont {L.~H.}\ \bibnamefont {Liu}},\ }\bibfield
  {title} {\bibinfo {title} {Heat diffusion in nanoparticle systems via
  near-field thermal photons},\ }\href
  {https://doi.org/https://doi.org/10.1016/j.ijheatmasstransfer.2022.123544}
  {\bibfield  {journal} {\bibinfo  {journal} {Int. J. Heat Mass Transf.}\
  }\textbf {\bibinfo {volume} {200}},\ \bibinfo {pages} {123544} (\bibinfo
  {year} {2023}{\natexlab{a}})}\BibitemShut {NoStop}%
\bibitem [{\citenamefont {Biehs}\ \emph {et~al.}(2011)\citenamefont {Biehs},
  \citenamefont {Rosa},\ and\ \citenamefont
  {Ben-Abdallah}}]{Biehs2011gratings}%
  \BibitemOpen
  \bibfield  {author} {\bibinfo {author} {\bibfnamefont {S.-A.}\ \bibnamefont
  {Biehs}}, \bibinfo {author} {\bibfnamefont {F.~S.~S.}\ \bibnamefont {Rosa}},\
  and\ \bibinfo {author} {\bibfnamefont {P.}~\bibnamefont {Ben-Abdallah}},\
  }\bibfield  {title} {\bibinfo {title} {Modulation of near-field heat transfer
  between two gratings},\ }\href {https://doi.org/10.1063/1.3596707} {\bibfield
   {journal} {\bibinfo  {journal} {Appl. Phys. Lett.}\ }\textbf {\bibinfo
  {volume} {98}},\ \bibinfo {pages} {243102} (\bibinfo {year}
  {2011})}\BibitemShut {NoStop}%
\bibitem [{\citenamefont {Kan}\ \emph {et~al.}(2019)\citenamefont {Kan},
  \citenamefont {Zhao},\ and\ \citenamefont {Zhang}}]{Kan2019prb}%
  \BibitemOpen
  \bibfield  {author} {\bibinfo {author} {\bibfnamefont {Y.~H.}\ \bibnamefont
  {Kan}}, \bibinfo {author} {\bibfnamefont {C.~Y.}\ \bibnamefont {Zhao}},\ and\
  \bibinfo {author} {\bibfnamefont {Z.~M.}\ \bibnamefont {Zhang}},\ }\bibfield
  {title} {\bibinfo {title} {Near-field radiative heat transfer in three-body
  systems with periodic structures},\ }\href
  {https://doi.org/10.1103/PhysRevB.99.035433} {\bibfield  {journal} {\bibinfo
  {journal} {Phys. Rev. B}\ }\textbf {\bibinfo {volume} {99}},\ \bibinfo
  {pages} {035433} (\bibinfo {year} {2019})}\BibitemShut {NoStop}%
\bibitem [{\citenamefont {Liu}\ \emph {et~al.}(2022{\natexlab{a}})\citenamefont
  {Liu}, \citenamefont {Chen}, \citenamefont {Caratenuto}, \citenamefont
  {Tian}, \citenamefont {Liu}, \citenamefont {Zhao},\ and\ \citenamefont
  {Zheng}}]{Zheng2022Materials_EMA}%
  \BibitemOpen
  \bibfield  {author} {\bibinfo {author} {\bibfnamefont {Y.}~\bibnamefont
  {Liu}}, \bibinfo {author} {\bibfnamefont {F.~Q.}\ \bibnamefont {Chen}},
  \bibinfo {author} {\bibfnamefont {A.}~\bibnamefont {Caratenuto}}, \bibinfo
  {author} {\bibfnamefont {Y.~P.}\ \bibnamefont {Tian}}, \bibinfo {author}
  {\bibfnamefont {X.~J.}\ \bibnamefont {Liu}}, \bibinfo {author} {\bibfnamefont
  {Y.~T.}\ \bibnamefont {Zhao}},\ and\ \bibinfo {author} {\bibfnamefont
  {Y.}~\bibnamefont {Zheng}},\ }\bibfield  {title} {\bibinfo {title} {Effective
  approximation method for nanogratings-induced near-field radiative heat
  transfer},\ }\href {https://doi.org/10.3390/ma15030998} {\bibfield  {journal}
  {\bibinfo  {journal} {Materials}\ }\textbf {\bibinfo {volume} {15}},\
  \bibinfo {pages} {998} (\bibinfo {year} {2022}{\natexlab{a}})}\BibitemShut
  {NoStop}%
\bibitem [{\citenamefont {Shen}\ \emph {et~al.}(2009)\citenamefont {Shen},
  \citenamefont {Narayanaswamy},\ and\ \citenamefont {Chen}}]{Shen2009}%
  \BibitemOpen
  \bibfield  {author} {\bibinfo {author} {\bibfnamefont {S.}~\bibnamefont
  {Shen}}, \bibinfo {author} {\bibfnamefont {A.}~\bibnamefont
  {Narayanaswamy}},\ and\ \bibinfo {author} {\bibfnamefont {G.}~\bibnamefont
  {Chen}},\ }\bibfield  {title} {\bibinfo {title} {Surface phonon polaritons
  mediated energy transfer between nanoscale gaps},\ }\href
  {https://doi.org/10.1021/nl901208v} {\bibfield  {journal} {\bibinfo
  {journal} {Nano Lett.}\ }\textbf {\bibinfo {volume} {9}},\ \bibinfo {pages}
  {2909} (\bibinfo {year} {2009})}\BibitemShut {NoStop}%
\bibitem [{\citenamefont {Rousseau}\ \emph {et~al.}(2009)\citenamefont
  {Rousseau}, \citenamefont {Siria}, \citenamefont {Jourdan}, \citenamefont
  {Volz}, \citenamefont {Comin}, \citenamefont {Chevrier},\ and\ \citenamefont
  {Greffet}}]{Rousseau2009}%
  \BibitemOpen
  \bibfield  {author} {\bibinfo {author} {\bibfnamefont {E.}~\bibnamefont
  {Rousseau}}, \bibinfo {author} {\bibfnamefont {A.}~\bibnamefont {Siria}},
  \bibinfo {author} {\bibfnamefont {G.}~\bibnamefont {Jourdan}}, \bibinfo
  {author} {\bibfnamefont {S.}~\bibnamefont {Volz}}, \bibinfo {author}
  {\bibfnamefont {F.}~\bibnamefont {Comin}}, \bibinfo {author} {\bibfnamefont
  {J.}~\bibnamefont {Chevrier}},\ and\ \bibinfo {author} {\bibfnamefont
  {J.-J.}\ \bibnamefont {Greffet}},\ }\bibfield  {title} {\bibinfo {title}
  {Radiative heat transfer at the nanoscale},\ }\href
  {https://doi.org/http://www.nature.com/nphoton/journal/v3/n9/suppinfo/nphoton.2009.144_S1.html}
  {\bibfield  {journal} {\bibinfo  {journal} {Nat. Photonics}\ }\textbf
  {\bibinfo {volume} {3}},\ \bibinfo {pages} {514} (\bibinfo {year}
  {2009})}\BibitemShut {NoStop}%
\bibitem [{\citenamefont {Ottens}\ \emph {et~al.}(2011)\citenamefont {Ottens},
  \citenamefont {Quetschke}, \citenamefont {Wise}, \citenamefont {Alemi},
  \citenamefont {Lundock}, \citenamefont {Mueller}, \citenamefont {Reitze},
  \citenamefont {Tanner},\ and\ \citenamefont {Whiting}}]{Ottens2011}%
  \BibitemOpen
  \bibfield  {author} {\bibinfo {author} {\bibfnamefont {R.~S.}\ \bibnamefont
  {Ottens}}, \bibinfo {author} {\bibfnamefont {V.}~\bibnamefont {Quetschke}},
  \bibinfo {author} {\bibfnamefont {S.}~\bibnamefont {Wise}}, \bibinfo {author}
  {\bibfnamefont {A.~A.}\ \bibnamefont {Alemi}}, \bibinfo {author}
  {\bibfnamefont {R.}~\bibnamefont {Lundock}}, \bibinfo {author} {\bibfnamefont
  {G.}~\bibnamefont {Mueller}}, \bibinfo {author} {\bibfnamefont {D.~H.}\
  \bibnamefont {Reitze}}, \bibinfo {author} {\bibfnamefont {D.~B.}\
  \bibnamefont {Tanner}},\ and\ \bibinfo {author} {\bibfnamefont {B.~F.}\
  \bibnamefont {Whiting}},\ }\bibfield  {title} {\bibinfo {title} {Near-field
  radiative heat transfer between macroscopic planar surfaces},\ }\href
  {https://doi.org/10.1103/PhysRevLett.107.014301} {\bibfield  {journal}
  {\bibinfo  {journal} {Phys. Rev. Lett.}\ }\textbf {\bibinfo {volume} {107}},\
  \bibinfo {pages} {014301} (\bibinfo {year} {2011})}\BibitemShut {NoStop}%
\bibitem [{\citenamefont {Song}\ \emph {et~al.}(2015)\citenamefont {Song},
  \citenamefont {Ganjeh}, \citenamefont {Sadat}, \citenamefont {Thompson},
  \citenamefont {Fiorino}, \citenamefont {Fernández-Hurtado}, \citenamefont
  {Feist}, \citenamefont {Garcia-Vidal}, \citenamefont {Cuevas}, \citenamefont
  {Reddy},\ and\ \citenamefont {Meyhofer}}]{Song2015}%
  \BibitemOpen
  \bibfield  {author} {\bibinfo {author} {\bibfnamefont {B.}~\bibnamefont
  {Song}}, \bibinfo {author} {\bibfnamefont {Y.}~\bibnamefont {Ganjeh}},
  \bibinfo {author} {\bibfnamefont {S.}~\bibnamefont {Sadat}}, \bibinfo
  {author} {\bibfnamefont {D.}~\bibnamefont {Thompson}}, \bibinfo {author}
  {\bibfnamefont {A.}~\bibnamefont {Fiorino}}, \bibinfo {author} {\bibfnamefont
  {V.}~\bibnamefont {Fernández-Hurtado}}, \bibinfo {author} {\bibfnamefont
  {J.}~\bibnamefont {Feist}}, \bibinfo {author} {\bibfnamefont {F.~J.}\
  \bibnamefont {Garcia-Vidal}}, \bibinfo {author} {\bibfnamefont {J.~C.}\
  \bibnamefont {Cuevas}}, \bibinfo {author} {\bibfnamefont {P.}~\bibnamefont
  {Reddy}},\ and\ \bibinfo {author} {\bibfnamefont {E.}~\bibnamefont
  {Meyhofer}},\ }\bibfield  {title} {\bibinfo {title} {Enhancement of
  near-field radiative heat transfer using polar dielectric thin films},\
  }\href {https://doi.org/10.1038/nnano.2015.6
  https://www.nature.com/articles/nnano.2015.6#supplementary-information}
  {\bibfield  {journal} {\bibinfo  {journal} {Nat. Nanotechnol.}\ }\textbf
  {\bibinfo {volume} {10}},\ \bibinfo {pages} {253} (\bibinfo {year}
  {2015})}\BibitemShut {NoStop}%
\bibitem [{\citenamefont {Lim}\ \emph {et~al.}(2015)\citenamefont {Lim},
  \citenamefont {Lee},\ and\ \citenamefont {Lee}}]{Lim2015}%
  \BibitemOpen
  \bibfield  {author} {\bibinfo {author} {\bibfnamefont {M.}~\bibnamefont
  {Lim}}, \bibinfo {author} {\bibfnamefont {S.~S.}\ \bibnamefont {Lee}},\ and\
  \bibinfo {author} {\bibfnamefont {B.~J.}\ \bibnamefont {Lee}},\ }\bibfield
  {title} {\bibinfo {title} {Near-field thermal radiation between doped silicon
  plates at nanoscale gaps},\ }\href
  {https://doi.org/10.1103/PhysRevB.91.195136} {\bibfield  {journal} {\bibinfo
  {journal} {Phys. Rev. B}\ }\textbf {\bibinfo {volume} {91}},\ \bibinfo
  {pages} {195136} (\bibinfo {year} {2015})}\BibitemShut {NoStop}%
\bibitem [{\citenamefont {Watjen}\ \emph {et~al.}(2016)\citenamefont {Watjen},
  \citenamefont {Zhao},\ and\ \citenamefont {Zhang}}]{Watjen2016}%
  \BibitemOpen
  \bibfield  {author} {\bibinfo {author} {\bibfnamefont {J.~I.}\ \bibnamefont
  {Watjen}}, \bibinfo {author} {\bibfnamefont {B.}~\bibnamefont {Zhao}},\ and\
  \bibinfo {author} {\bibfnamefont {Z.~M.}\ \bibnamefont {Zhang}},\ }\bibfield
  {title} {\bibinfo {title} {Near-field radiative heat transfer between
  doped-si parallel plates separated by a spacing down to 200 nm},\ }\href
  {https://doi.org/10.1063/1.4967384} {\bibfield  {journal} {\bibinfo
  {journal} {Appl. Phys. Lett.}\ }\textbf {\bibinfo {volume} {109}} (\bibinfo
  {year} {2016})}\BibitemShut {NoStop}%
\bibitem [{\citenamefont {Ghashami}\ \emph {et~al.}(2018)\citenamefont
  {Ghashami}, \citenamefont {Geng}, \citenamefont {Kim}, \citenamefont
  {Iacopino}, \citenamefont {Cho},\ and\ \citenamefont {Park}}]{Ghashami2018}%
  \BibitemOpen
  \bibfield  {author} {\bibinfo {author} {\bibfnamefont {M.}~\bibnamefont
  {Ghashami}}, \bibinfo {author} {\bibfnamefont {H.}~\bibnamefont {Geng}},
  \bibinfo {author} {\bibfnamefont {T.}~\bibnamefont {Kim}}, \bibinfo {author}
  {\bibfnamefont {N.}~\bibnamefont {Iacopino}}, \bibinfo {author}
  {\bibfnamefont {S.~K.}\ \bibnamefont {Cho}},\ and\ \bibinfo {author}
  {\bibfnamefont {K.}~\bibnamefont {Park}},\ }\bibfield  {title} {\bibinfo
  {title} {Precision measurement of phonon-polaritonic near-field energy
  transfer between macroscale planar structures under large thermal
  gradients},\ }\href {https://doi.org/10.1103/PhysRevLett.120.175901}
  {\bibfield  {journal} {\bibinfo  {journal} {Phys. Rev. Lett.}\ }\textbf
  {\bibinfo {volume} {120}},\ \bibinfo {pages} {175901} (\bibinfo {year}
  {2018})}\BibitemShut {NoStop}%
\bibitem [{\citenamefont {Yang}\ \emph {et~al.}(2018)\citenamefont {Yang},
  \citenamefont {Du}, \citenamefont {Su} \emph {et~al.}}]{Yungui2018NC}%
  \BibitemOpen
  \bibfield  {author} {\bibinfo {author} {\bibfnamefont {J.}~\bibnamefont
  {Yang}}, \bibinfo {author} {\bibfnamefont {W.}~\bibnamefont {Du}}, \bibinfo
  {author} {\bibfnamefont {Y.~S.}\ \bibnamefont {Su}}, \emph {et~al.},\
  }\bibfield  {title} {\bibinfo {title} {Observing of the super-planckian
  near-field thermal radiation between graphene sheets},\ }\href
  {https://doi.org/10.1038/s41467-018-06163-8} {\bibfield  {journal} {\bibinfo
  {journal} {Nat. Commun.}\ }\textbf {\bibinfo {volume} {9}},\ \bibinfo {pages}
  {4033} (\bibinfo {year} {2018})}\BibitemShut {NoStop}%
\bibitem [{\citenamefont {DeSutter}\ \emph {et~al.}(2019)\citenamefont
  {DeSutter}, \citenamefont {Tang},\ and\ \citenamefont
  {Francoeur}}]{DeSutter2019}%
  \BibitemOpen
  \bibfield  {author} {\bibinfo {author} {\bibfnamefont {J.}~\bibnamefont
  {DeSutter}}, \bibinfo {author} {\bibfnamefont {L.}~\bibnamefont {Tang}},\
  and\ \bibinfo {author} {\bibfnamefont {M.}~\bibnamefont {Francoeur}},\
  }\bibfield  {title} {\bibinfo {title} {A near-field radiative heat transfer
  device},\ }\href {https://doi.org/10.1038/s41565-019-0483-1} {\bibfield
  {journal} {\bibinfo  {journal} {Nat. Nanotechnol.}\ }\textbf {\bibinfo
  {volume} {14}},\ \bibinfo {pages} {751} (\bibinfo {year} {2019})}\BibitemShut
  {NoStop}%
\bibitem [{\citenamefont {Iqbal}\ \emph {et~al.}(2023)\citenamefont {Iqbal},
  \citenamefont {Zhang}, \citenamefont {Wang} \emph
  {et~al.}}]{Yungui2023PRAppl}%
  \BibitemOpen
  \bibfield  {author} {\bibinfo {author} {\bibfnamefont {N.}~\bibnamefont
  {Iqbal}}, \bibinfo {author} {\bibfnamefont {S.}~\bibnamefont {Zhang}},
  \bibinfo {author} {\bibfnamefont {S.}~\bibnamefont {Wang}}, \emph {et~al.},\
  }\bibfield  {title} {\bibinfo {title} {Measuring near-field radiative heat
  transfer in a graphene-$\mathrm{Si}\mathrm{C}$ heterostructure},\ }\href
  {https://doi.org/10.1103/PhysRevApplied.19.024019} {\bibfield  {journal}
  {\bibinfo  {journal} {Phys. Rev. Appl.}\ }\textbf {\bibinfo {volume} {19}},\
  \bibinfo {pages} {024019} (\bibinfo {year} {2023})}\BibitemShut {NoStop}%
\bibitem [{\citenamefont {Yang}\ and\ \citenamefont
  {Wang}(2016)}]{Yang2017prl}%
  \BibitemOpen
  \bibfield  {author} {\bibinfo {author} {\bibfnamefont {Y.}~\bibnamefont
  {Yang}}\ and\ \bibinfo {author} {\bibfnamefont {L.}~\bibnamefont {Wang}},\
  }\bibfield  {title} {\bibinfo {title} {Spectrally enhancing near-field
  radiative transfer between metallic gratings by exciting magnetic polaritons
  in nanometric vacuum gaps},\ }\href
  {https://doi.org/10.1103/PhysRevLett.117.044301} {\bibfield  {journal}
  {\bibinfo  {journal} {Phys. Rev. Lett.}\ }\textbf {\bibinfo {volume} {117}},\
  \bibinfo {pages} {044301} (\bibinfo {year} {2016})}\BibitemShut {NoStop}%
\bibitem [{\citenamefont {Messina}\ \emph {et~al.}(2017)\citenamefont
  {Messina}, \citenamefont {Noto}, \citenamefont {Guizal},\ and\ \citenamefont
  {Antezza}}]{Messina2017PRB}%
  \BibitemOpen
  \bibfield  {author} {\bibinfo {author} {\bibfnamefont {R.}~\bibnamefont
  {Messina}}, \bibinfo {author} {\bibfnamefont {A.}~\bibnamefont {Noto}},
  \bibinfo {author} {\bibfnamefont {B.}~\bibnamefont {Guizal}},\ and\ \bibinfo
  {author} {\bibfnamefont {M.}~\bibnamefont {Antezza}},\ }\bibfield  {title}
  {\bibinfo {title} {Radiative heat transfer between metallic gratings using
  fourier modal method with adaptive spatial resolution},\ }\href
  {https://doi.org/10.1103/PhysRevB.95.125404} {\bibfield  {journal} {\bibinfo
  {journal} {Phys. Rev. B}\ }\textbf {\bibinfo {volume} {95}},\ \bibinfo
  {pages} {125404} (\bibinfo {year} {2017})}\BibitemShut {NoStop}%
\bibitem [{\citenamefont {Cui}\ \emph {et~al.}(2023)\citenamefont {Cui},
  \citenamefont {Zhou}, \citenamefont {Zhang},\ and\ \citenamefont
  {Yi}}]{Hongliang_JHMT}%
  \BibitemOpen
  \bibfield  {author} {\bibinfo {author} {\bibfnamefont {G.-C.}\ \bibnamefont
  {Cui}}, \bibinfo {author} {\bibfnamefont {C.-L.}\ \bibnamefont {Zhou}},
  \bibinfo {author} {\bibfnamefont {Y.}~\bibnamefont {Zhang}},\ and\ \bibinfo
  {author} {\bibfnamefont {H.-L.}\ \bibnamefont {Yi}},\ }\bibfield  {title}
  {\bibinfo {title} {{Significant Enhancement of Near-Field Radiative Heat
  Transfer by Misaligned Bilayer Heterostructure of Graphene-Covered
  Gratings}},\ }\href {https://doi.org/10.1115/1.4063644} {\bibfield  {journal}
  {\bibinfo  {journal} {ASME J. Heat Mass Transfer}\ }\textbf {\bibinfo
  {volume} {146}},\ \bibinfo {pages} {022801} (\bibinfo {year}
  {2023})}\BibitemShut {NoStop}%
\bibitem [{\citenamefont {Svetovoy}\ \emph {et~al.}(2012)\citenamefont
  {Svetovoy}, \citenamefont {van Zwol},\ and\ \citenamefont
  {Chevrier}}]{Svetovoy2012prb}%
  \BibitemOpen
  \bibfield  {author} {\bibinfo {author} {\bibfnamefont {V.~B.}\ \bibnamefont
  {Svetovoy}}, \bibinfo {author} {\bibfnamefont {P.~J.}\ \bibnamefont {van
  Zwol}},\ and\ \bibinfo {author} {\bibfnamefont {J.}~\bibnamefont
  {Chevrier}},\ }\bibfield  {title} {\bibinfo {title} {Plasmon enhanced
  near-field radiative heat transfer for graphene covered dielectrics},\ }\href
  {https://doi.org/10.1103/PhysRevB.85.155418} {\bibfield  {journal} {\bibinfo
  {journal} {Phys. Rev. B}\ }\textbf {\bibinfo {volume} {85}},\ \bibinfo
  {pages} {155418} (\bibinfo {year} {2012})}\BibitemShut {NoStop}%
\bibitem [{\citenamefont {Ilic}\ \emph {et~al.}(2012)\citenamefont {Ilic},
  \citenamefont {Jablan}, \citenamefont {Joannopoulos}, \citenamefont
  {Celanovic}, \citenamefont {Buljan},\ and\ \citenamefont {Solja\ifmmode
  \check{c}\else \v{c}\fi{}i\ifmmode~\acute{c}\else
  \'{c}\fi{}}}]{Ognjen2012prb}%
  \BibitemOpen
  \bibfield  {author} {\bibinfo {author} {\bibfnamefont {O.}~\bibnamefont
  {Ilic}}, \bibinfo {author} {\bibfnamefont {M.}~\bibnamefont {Jablan}},
  \bibinfo {author} {\bibfnamefont {J.~D.}\ \bibnamefont {Joannopoulos}},
  \bibinfo {author} {\bibfnamefont {I.}~\bibnamefont {Celanovic}}, \bibinfo
  {author} {\bibfnamefont {H.}~\bibnamefont {Buljan}},\ and\ \bibinfo {author}
  {\bibfnamefont {M.}~\bibnamefont {Solja\ifmmode \check{c}\else
  \v{c}\fi{}i\ifmmode~\acute{c}\else \'{c}\fi{}}},\ }\bibfield  {title}
  {\bibinfo {title} {Near-field thermal radiation transfer controlled by
  plasmons in graphene},\ }\href {https://doi.org/10.1103/PhysRevB.85.155422}
  {\bibfield  {journal} {\bibinfo  {journal} {Phys. Rev. B}\ }\textbf {\bibinfo
  {volume} {85}},\ \bibinfo {pages} {155422} (\bibinfo {year}
  {2012})}\BibitemShut {NoStop}%
\bibitem [{\citenamefont {Zheng}\ \emph {et~al.}(2017)\citenamefont {Zheng},
  \citenamefont {Liu}, \citenamefont {Wang},\ and\ \citenamefont
  {Xuan}}]{Zheng2017}%
  \BibitemOpen
  \bibfield  {author} {\bibinfo {author} {\bibfnamefont {Z.~H.}\ \bibnamefont
  {Zheng}}, \bibinfo {author} {\bibfnamefont {X.~L.}\ \bibnamefont {Liu}},
  \bibinfo {author} {\bibfnamefont {A.}~\bibnamefont {Wang}},\ and\ \bibinfo
  {author} {\bibfnamefont {Y.~M.}\ \bibnamefont {Xuan}},\ }\bibfield  {title}
  {\bibinfo {title} {Graphene-assisted near-field radiative thermal rectifier
  based on phase transition of vanadium dioxide ({VO}$_2$)},\ }\href
  {https://doi.org/https://doi.org/10.1016/j.ijheatmasstransfer.2017.01.107}
  {\bibfield  {journal} {\bibinfo  {journal} {Int. J. Heat Mass Transf.}\
  }\textbf {\bibinfo {volume} {109}},\ \bibinfo {pages} {63 } (\bibinfo {year}
  {2017})}\BibitemShut {NoStop}%
\bibitem [{\citenamefont {Volokitin}(2017)}]{Volokitin2017Dey}%
  \BibitemOpen
  \bibfield  {author} {\bibinfo {author} {\bibfnamefont {A.}~\bibnamefont
  {Volokitin}},\ }\bibfield  {title} {\bibinfo {title} {Casimir friction and
  near-field radiative heat transfer in graphene structures},\ }\href
  {https://doi.org/doi:10.1515/zna-2016-0367} {\bibfield  {journal} {\bibinfo
  {journal} {Z. Naturforsch. A}\ }\textbf {\bibinfo {volume} {72}},\ \bibinfo
  {pages} {171} (\bibinfo {year} {2017})}\BibitemShut {NoStop}%
\bibitem [{\citenamefont {Zhao}\ \emph {et~al.}(2017)\citenamefont {Zhao},
  \citenamefont {Guizal}, \citenamefont {Zhang}, \citenamefont {Fan},\ and\
  \citenamefont {Antezza}}]{Zhao2017prb}%
  \BibitemOpen
  \bibfield  {author} {\bibinfo {author} {\bibfnamefont {B.}~\bibnamefont
  {Zhao}}, \bibinfo {author} {\bibfnamefont {B.}~\bibnamefont {Guizal}},
  \bibinfo {author} {\bibfnamefont {Z.~M.}\ \bibnamefont {Zhang}}, \bibinfo
  {author} {\bibfnamefont {S.}~\bibnamefont {Fan}},\ and\ \bibinfo {author}
  {\bibfnamefont {M.}~\bibnamefont {Antezza}},\ }\bibfield  {title} {\bibinfo
  {title} {Near-field heat transfer between graphene/{hBN} multilayers},\
  }\href {https://doi.org/10.1103/PhysRevB.95.245437} {\bibfield  {journal}
  {\bibinfo  {journal} {Phys. Rev. B}\ }\textbf {\bibinfo {volume} {95}},\
  \bibinfo {pages} {245437} (\bibinfo {year} {2017})}\BibitemShut {NoStop}%
\bibitem [{\citenamefont {Shi}\ \emph {et~al.}(2021)\citenamefont {Shi},
  \citenamefont {Chen}, \citenamefont {Xu}, \citenamefont {Evans},\ and\
  \citenamefont {He}}]{Shi2021am}%
  \BibitemOpen
  \bibfield  {author} {\bibinfo {author} {\bibfnamefont {K.~Z.}\ \bibnamefont
  {Shi}}, \bibinfo {author} {\bibfnamefont {Z.~Y.}\ \bibnamefont {Chen}},
  \bibinfo {author} {\bibfnamefont {X.}~\bibnamefont {Xu}}, \bibinfo {author}
  {\bibfnamefont {J.}~\bibnamefont {Evans}},\ and\ \bibinfo {author}
  {\bibfnamefont {S.~L.}\ \bibnamefont {He}},\ }\bibfield  {title} {\bibinfo
  {title} {Optimized colossal near-field thermal radiation enabled by
  manipulating coupled plasmon polariton geometry},\ }\href
  {https://doi.org/https://doi.org/10.1002/adma.202106097} {\bibfield
  {journal} {\bibinfo  {journal} {Advanced Materials}\ }\textbf {\bibinfo
  {volume} {33}},\ \bibinfo {pages} {2106097} (\bibinfo {year}
  {2021})}\BibitemShut {NoStop}%
\bibitem [{\citenamefont {Liu}\ \emph {et~al.}(2022{\natexlab{b}})\citenamefont
  {Liu}, \citenamefont {Ge}, \citenamefont {Yu}, \citenamefont {Cui},\ and\
  \citenamefont {Wu}}]{LiuES2022}%
  \BibitemOpen
  \bibfield  {author} {\bibinfo {author} {\bibfnamefont {R.~Y.}\ \bibnamefont
  {Liu}}, \bibinfo {author} {\bibfnamefont {L.~X.}\ \bibnamefont {Ge}},
  \bibinfo {author} {\bibfnamefont {H.~Y.}\ \bibnamefont {Yu}}, \bibinfo
  {author} {\bibfnamefont {Z.}~\bibnamefont {Cui}},\ and\ \bibinfo {author}
  {\bibfnamefont {X.~H.}\ \bibnamefont {Wu}},\ }\bibfield  {title} {\bibinfo
  {title} {Near-field radiative heat transfer via coupling graphene plasmons
  with different phonon polaritons in the reststrahlen bands},\ }\href
  {https://doi.org/10.30919/es8d529} {\bibfield  {journal} {\bibinfo  {journal}
  {Eng. Sci.}\ }\textbf {\bibinfo {volume} {18}},\ \bibinfo {pages} {224}
  (\bibinfo {year} {2022}{\natexlab{b}})}\BibitemShut {NoStop}%
\bibitem [{\citenamefont {Lu}\ \emph {et~al.}(2022)\citenamefont {Lu},
  \citenamefont {Zhang}, \citenamefont {Ou}, \citenamefont {Li}, \citenamefont
  {Zhou}, \citenamefont {Song}, \citenamefont {Luo},\ and\ \citenamefont
  {Cheng}}]{Lu2022small}%
  \BibitemOpen
  \bibfield  {author} {\bibinfo {author} {\bibfnamefont {L.}~\bibnamefont
  {Lu}}, \bibinfo {author} {\bibfnamefont {B.}~\bibnamefont {Zhang}}, \bibinfo
  {author} {\bibfnamefont {H.}~\bibnamefont {Ou}}, \bibinfo {author}
  {\bibfnamefont {B.~W.}\ \bibnamefont {Li}}, \bibinfo {author} {\bibfnamefont
  {K.}~\bibnamefont {Zhou}}, \bibinfo {author} {\bibfnamefont {J.~L.}\
  \bibnamefont {Song}}, \bibinfo {author} {\bibfnamefont {Z.~X.}\ \bibnamefont
  {Luo}},\ and\ \bibinfo {author} {\bibfnamefont {Q.}~\bibnamefont {Cheng}},\
  }\bibfield  {title} {\bibinfo {title} {Enhanced near-field radiative heat
  transfer between graphene/{hBN} systems},\ }\href
  {https://doi.org/https://doi.org/10.1002/smll.202108032} {\bibfield
  {journal} {\bibinfo  {journal} {Small}\ }\textbf {\bibinfo {volume} {18}},\
  \bibinfo {pages} {2108032} (\bibinfo {year} {2022})}\BibitemShut {NoStop}%
\bibitem [{\citenamefont {Wang}\ and\ \citenamefont {Antezza}(2024)}]{JSWM}%
  \BibitemOpen
  \bibfield  {author} {\bibinfo {author} {\bibfnamefont {J.-S.}\ \bibnamefont
  {Wang}}\ and\ \bibinfo {author} {\bibfnamefont {M.}~\bibnamefont {Antezza}},\
  }\bibfield  {title} {\bibinfo {title} {Photon mediated energy, linear and
  angular momentum transport in fullerene and graphene systems beyond local
  equilibrium},\ }\href {https://doi.org/10.1103/PhysRevB.109.125105}
  {\bibfield  {journal} {\bibinfo  {journal} {Phys. Rev. B}\ }\textbf {\bibinfo
  {volume} {109}},\ \bibinfo {pages} {125105} (\bibinfo {year}
  {2024})}\BibitemShut {NoStop}%
\bibitem [{\citenamefont {Zhang}\ \emph {et~al.}(2022)\citenamefont {Zhang},
  \citenamefont {Antezza},\ and\ \citenamefont {Wang}}]{GrafBilMA}%
  \BibitemOpen
  \bibfield  {author} {\bibinfo {author} {\bibfnamefont {Y.-M.}\ \bibnamefont
  {Zhang}}, \bibinfo {author} {\bibfnamefont {M.}~\bibnamefont {Antezza}},\
  and\ \bibinfo {author} {\bibfnamefont {J.-S.}\ \bibnamefont {Wang}},\
  }\bibfield  {title} {\bibinfo {title} {Controllable thermal radiation from
  twisted bilayer graphene},\ }\href
  {https://doi.org/https://doi.org/10.1016/j.ijheatmasstransfer.2022.123076}
  {\bibfield  {journal} {\bibinfo  {journal} {Int. J Heat Mass Transf.}\
  }\textbf {\bibinfo {volume} {194}},\ \bibinfo {pages} {123076} (\bibinfo
  {year} {2022})}\BibitemShut {NoStop}%
\bibitem [{\citenamefont {Liu}\ and\ \citenamefont {Zhang}(2015)}]{Liu2015APL}%
  \BibitemOpen
  \bibfield  {author} {\bibinfo {author} {\bibfnamefont {X.~L.}\ \bibnamefont
  {Liu}}\ and\ \bibinfo {author} {\bibfnamefont {Z.~M.}\ \bibnamefont
  {Zhang}},\ }\bibfield  {title} {\bibinfo {title} {{Giant enhancement of
  nanoscale thermal radiation based on hyperbolic graphene plasmons}},\ }\href
  {https://doi.org/10.1063/1.4932958} {\bibfield  {journal} {\bibinfo
  {journal} {Appl. Phys. Lett.}\ }\textbf {\bibinfo {volume} {107}},\ \bibinfo
  {pages} {143114} (\bibinfo {year} {2015})}\BibitemShut {NoStop}%
\bibitem [{\citenamefont {Hu}\ \emph {et~al.}(2020)\citenamefont {Hu},
  \citenamefont {Li}, \citenamefont {Zhu},\ and\ \citenamefont
  {Yang}}]{Yang2020prAppl}%
  \BibitemOpen
  \bibfield  {author} {\bibinfo {author} {\bibfnamefont {Y.~Z.}\ \bibnamefont
  {Hu}}, \bibinfo {author} {\bibfnamefont {H.}~\bibnamefont {Li}}, \bibinfo
  {author} {\bibfnamefont {Y.~G.}\ \bibnamefont {Zhu}},\ and\ \bibinfo {author}
  {\bibfnamefont {Y.}~\bibnamefont {Yang}},\ }\bibfield  {title} {\bibinfo
  {title} {Enhanced near-field radiative heat transport between graphene
  metasurfaces with symmetric nanopatterns},\ }\href
  {https://doi.org/10.1103/PhysRevApplied.14.044054} {\bibfield  {journal}
  {\bibinfo  {journal} {Phys. Rev. Appl.}\ }\textbf {\bibinfo {volume} {14}},\
  \bibinfo {pages} {044054} (\bibinfo {year} {2020})}\BibitemShut {NoStop}%
\bibitem [{\citenamefont {Luo}\ \emph {et~al.}(2023{\natexlab{b}})\citenamefont
  {Luo}, \citenamefont {Jeyar}, \citenamefont {Guizal}, \citenamefont {Zhao},\
  and\ \citenamefont {Antezza}}]{Luo2023apl_NFRHT_grating}%
  \BibitemOpen
  \bibfield  {author} {\bibinfo {author} {\bibfnamefont {M.~G.}\ \bibnamefont
  {Luo}}, \bibinfo {author} {\bibfnamefont {Y.}~\bibnamefont {Jeyar}}, \bibinfo
  {author} {\bibfnamefont {B.}~\bibnamefont {Guizal}}, \bibinfo {author}
  {\bibfnamefont {J.~M.}\ \bibnamefont {Zhao}},\ and\ \bibinfo {author}
  {\bibfnamefont {M.}~\bibnamefont {Antezza}},\ }\bibfield  {title} {\bibinfo
  {title} {{Effect of graphene grating coating on near-field radiative heat
  transfer}},\ }\href {https://doi.org/10.1063/5.0182725} {\bibfield  {journal}
  {\bibinfo  {journal} {Appl. Phys. Lett.}\ }\textbf {\bibinfo {volume}
  {123}},\ \bibinfo {pages} {253902} (\bibinfo {year}
  {2023}{\natexlab{b}})}\BibitemShut {NoStop}%
\bibitem [{\citenamefont {He}\ \emph {et~al.}(2020{\natexlab{a}})\citenamefont
  {He}, \citenamefont {Qi}, \citenamefont {Ren}, \citenamefont {Zhao},\ and\
  \citenamefont {Antezza}}]{He2020ol}%
  \BibitemOpen
  \bibfield  {author} {\bibinfo {author} {\bibfnamefont {M.~J.}\ \bibnamefont
  {He}}, \bibinfo {author} {\bibfnamefont {H.}~\bibnamefont {Qi}}, \bibinfo
  {author} {\bibfnamefont {Y.~T.}\ \bibnamefont {Ren}}, \bibinfo {author}
  {\bibfnamefont {Y.}~\bibnamefont {Zhao}},\ and\ \bibinfo {author}
  {\bibfnamefont {M.}~\bibnamefont {Antezza}},\ }\bibfield  {title} {\bibinfo
  {title} {Active control of near-field radiative heat transfer by a
  graphene-gratings coating-twisting method},\ }\href
  {https://doi.org/10.1364/OL.392371} {\bibfield  {journal} {\bibinfo
  {journal} {Opt. Lett.}\ }\textbf {\bibinfo {volume} {45}},\ \bibinfo {pages}
  {2914} (\bibinfo {year} {2020}{\natexlab{a}})}\BibitemShut {NoStop}%
\bibitem [{\citenamefont {He}\ \emph {et~al.}(2020{\natexlab{b}})\citenamefont
  {He}, \citenamefont {Qi}, \citenamefont {Ren}, \citenamefont {Zhao},\ and\
  \citenamefont {Antezza}}]{He2020ijhmt}%
  \BibitemOpen
  \bibfield  {author} {\bibinfo {author} {\bibfnamefont {M.~J.}\ \bibnamefont
  {He}}, \bibinfo {author} {\bibfnamefont {H.}~\bibnamefont {Qi}}, \bibinfo
  {author} {\bibfnamefont {Y.-T.}\ \bibnamefont {Ren}}, \bibinfo {author}
  {\bibfnamefont {Y.-J.}\ \bibnamefont {Zhao}},\ and\ \bibinfo {author}
  {\bibfnamefont {M.}~\bibnamefont {Antezza}},\ }\bibfield  {title} {\bibinfo
  {title} {Magnetoplasmonic manipulation of nanoscale thermal radiation using
  twisted graphene gratings},\ }\href
  {https://doi.org/https://doi.org/10.1016/j.ijheatmasstransfer.2020.119305}
  {\bibfield  {journal} {\bibinfo  {journal} {Int. J. Heat Mass Transf.}\
  }\textbf {\bibinfo {volume} {150}},\ \bibinfo {pages} {119305} (\bibinfo
  {year} {2020}{\natexlab{b}})}\BibitemShut {NoStop}%
\bibitem [{\citenamefont {Luo}\ \emph {et~al.}(2020{\natexlab{b}})\citenamefont
  {Luo}, \citenamefont {Zhao},\ and\ \citenamefont {Antezza}}]{Luo2020apl}%
  \BibitemOpen
  \bibfield  {author} {\bibinfo {author} {\bibfnamefont {M.~G.}\ \bibnamefont
  {Luo}}, \bibinfo {author} {\bibfnamefont {J.~M.}\ \bibnamefont {Zhao}},\ and\
  \bibinfo {author} {\bibfnamefont {M.}~\bibnamefont {Antezza}},\ }\bibfield
  {title} {\bibinfo {title} {{Near-field radiative heat transfer between
  twisted nanoparticle gratings}},\ }\href {https://doi.org/10.1063/5.0018329}
  {\bibfield  {journal} {\bibinfo  {journal} {Appl. Phys. Lett.}\ }\textbf
  {\bibinfo {volume} {117}},\ \bibinfo {pages} {053901} (\bibinfo {year}
  {2020}{\natexlab{b}})}\BibitemShut {NoStop}%
\bibitem [{\citenamefont {Zhou}\ \emph
  {et~al.}(2022{\natexlab{a}})\citenamefont {Zhou}, \citenamefont {Zhang},\
  and\ \citenamefont {Yi}}]{Zhou2022langmuir}%
  \BibitemOpen
  \bibfield  {author} {\bibinfo {author} {\bibfnamefont {C.-L.}\ \bibnamefont
  {Zhou}}, \bibinfo {author} {\bibfnamefont {Y.}~\bibnamefont {Zhang}},\ and\
  \bibinfo {author} {\bibfnamefont {H.-L.}\ \bibnamefont {Yi}},\ }\bibfield
  {title} {\bibinfo {title} {Enhancement and manipulation of near-field thermal
  radiation using hybrid hyperbolic polaritons},\ }\href
  {https://doi.org/10.1021/acs.langmuir.2c00467} {\bibfield  {journal}
  {\bibinfo  {journal} {Langmuir}\ }\textbf {\bibinfo {volume} {38}},\ \bibinfo
  {pages} {7689} (\bibinfo {year} {2022}{\natexlab{a}})}\BibitemShut {NoStop}%
\bibitem [{\citenamefont {Guizal}\ and\ \citenamefont
  {Felbacq}(1999)}]{Guizal}%
  \BibitemOpen
  \bibfield  {author} {\bibinfo {author} {\bibfnamefont {B.}~\bibnamefont
  {Guizal}}\ and\ \bibinfo {author} {\bibfnamefont {D.}~\bibnamefont
  {Felbacq}},\ }\bibfield  {title} {\bibinfo {title} {Electromagnetic beam
  diffraction by a finite strip grating},\ }\href
  {https://doi.org/10.1016/S0030-4018(99)00192-3} {\bibfield  {journal}
  {\bibinfo  {journal} {Optics Communications}\ }\textbf {\bibinfo {volume}
  {165}},\ \bibinfo {pages} {1} (\bibinfo {year} {1999})}\BibitemShut {NoStop}%
\bibitem [{\citenamefont {Hwang}(2020{\natexlab{a}})}]{Taiwan_LBF}%
  \BibitemOpen
  \bibfield  {author} {\bibinfo {author} {\bibfnamefont {R.-B.}\ \bibnamefont
  {Hwang}},\ }\bibfield  {title} {\bibinfo {title} {Highly improved convergence
  approach incorporating edge conditions for scattering analysis of graphene
  gratings},\ }\href {https://doi.org/10.1038/s41598-020-69827-w} {\bibfield
  {journal} {\bibinfo  {journal} {Scientific Reports}\ }\textbf {\bibinfo
  {volume} {10}},\ \bibinfo {pages} {12855} (\bibinfo {year}
  {2020}{\natexlab{a}})}\BibitemShut {NoStop}%
\bibitem [{\citenamefont {Jeyar}\ \emph
  {et~al.}(2023{\natexlab{a}})\citenamefont {Jeyar}, \citenamefont {Antezza},\
  and\ \citenamefont {Guizal}}]{PRE23}%
  \BibitemOpen
  \bibfield  {author} {\bibinfo {author} {\bibfnamefont {Y.}~\bibnamefont
  {Jeyar}}, \bibinfo {author} {\bibfnamefont {M.}~\bibnamefont {Antezza}},\
  and\ \bibinfo {author} {\bibfnamefont {B.}~\bibnamefont {Guizal}},\
  }\bibfield  {title} {\bibinfo {title} {Electromagnetic scattering by a
  partially graphene-coated dielectric cylinder: Efficient computation and
  multiple plasmonic resonances},\ }\href
  {https://doi.org/10.1103/PhysRevE.107.025306} {\bibfield  {journal} {\bibinfo
   {journal} {Phys. Rev. E}\ }\textbf {\bibinfo {volume} {107}},\ \bibinfo
  {pages} {025306} (\bibinfo {year} {2023}{\natexlab{a}})}\BibitemShut
  {NoStop}%
\bibitem [{\citenamefont {Jeyar}\ \emph
  {et~al.}(2023{\natexlab{b}})\citenamefont {Jeyar}, \citenamefont {Luo},
  \citenamefont {Austry}, \citenamefont {Guizal}, \citenamefont {Zheng},
  \citenamefont {Chan},\ and\ \citenamefont {Antezza}}]{Luo2023Casimir_gg}%
  \BibitemOpen
  \bibfield  {author} {\bibinfo {author} {\bibfnamefont {Y.}~\bibnamefont
  {Jeyar}}, \bibinfo {author} {\bibfnamefont {M.~G.}\ \bibnamefont {Luo}},
  \bibinfo {author} {\bibfnamefont {K.}~\bibnamefont {Austry}}, \bibinfo
  {author} {\bibfnamefont {B.}~\bibnamefont {Guizal}}, \bibinfo {author}
  {\bibfnamefont {Y.}~\bibnamefont {Zheng}}, \bibinfo {author} {\bibfnamefont
  {H.~B.}\ \bibnamefont {Chan}},\ and\ \bibinfo {author} {\bibfnamefont
  {M.}~\bibnamefont {Antezza}},\ }\bibfield  {title} {\bibinfo {title} {Tunable
  nonadditivity in the casimir-lifshitz force between graphene gratings},\
  }\href {https://doi.org/10.1103/PhysRevA.108.062811} {\bibfield  {journal}
  {\bibinfo  {journal} {Phys. Rev. A}\ }\textbf {\bibinfo {volume} {108}},\
  \bibinfo {pages} {062811} (\bibinfo {year} {2023}{\natexlab{b}})}\BibitemShut
  {NoStop}%
\bibitem [{\citenamefont {Falkovsky}\ and\ \citenamefont
  {Varlamov}(2007)}]{Falkovsky2007}%
  \BibitemOpen
  \bibfield  {author} {\bibinfo {author} {\bibfnamefont {L.~A.}\ \bibnamefont
  {Falkovsky}}\ and\ \bibinfo {author} {\bibfnamefont {A.~A.}\ \bibnamefont
  {Varlamov}},\ }\bibfield  {title} {\bibinfo {title} {Space-time dispersion of
  graphene conductivity},\ }\href {https://doi.org/10.1140/epjb/e2007-00142-3}
  {\bibfield  {journal} {\bibinfo  {journal} {Eur. Phys. J. B}\ }\textbf
  {\bibinfo {volume} {56}},\ \bibinfo {pages} {281} (\bibinfo {year}
  {2007})}\BibitemShut {NoStop}%
\bibitem [{\citenamefont {Falkovsky}(2008)}]{Falkovsky2008}%
  \BibitemOpen
  \bibfield  {author} {\bibinfo {author} {\bibfnamefont {L.~A.}\ \bibnamefont
  {Falkovsky}},\ }\bibfield  {title} {\bibinfo {title} {Optical properties of
  graphene},\ }\href {https://doi.org/10.1088/1742-6596/129/1/012004}
  {\bibfield  {journal} {\bibinfo  {journal} {J. Phys. Conf. Ser.}\ }\textbf
  {\bibinfo {volume} {129}},\ \bibinfo {pages} {012004} (\bibinfo {year}
  {2008})}\BibitemShut {NoStop}%
\bibitem [{\citenamefont {Awan}\ \emph {et~al.}(2016)\citenamefont {Awan},
  \citenamefont {Lombardo}, \citenamefont {Colli}, \citenamefont {Privitera},
  \citenamefont {Kulmala}, \citenamefont {Kivioja}, \citenamefont {Koshino},\
  and\ \citenamefont {Ferrari}}]{Awan_2016}%
  \BibitemOpen
  \bibfield  {author} {\bibinfo {author} {\bibfnamefont {S.~A.}\ \bibnamefont
  {Awan}}, \bibinfo {author} {\bibfnamefont {A.}~\bibnamefont {Lombardo}},
  \bibinfo {author} {\bibfnamefont {A.}~\bibnamefont {Colli}}, \bibinfo
  {author} {\bibfnamefont {G.}~\bibnamefont {Privitera}}, \bibinfo {author}
  {\bibfnamefont {T.~S.}\ \bibnamefont {Kulmala}}, \bibinfo {author}
  {\bibfnamefont {J.~M.}\ \bibnamefont {Kivioja}}, \bibinfo {author}
  {\bibfnamefont {M.}~\bibnamefont {Koshino}},\ and\ \bibinfo {author}
  {\bibfnamefont {A.~C.}\ \bibnamefont {Ferrari}},\ }\bibfield  {title}
  {\bibinfo {title} {Transport conductivity of graphene at {RF} and microwave
  frequencies},\ }\href {https://doi.org/10.1088/2053-1583/3/1/015010}
  {\bibfield  {journal} {\bibinfo  {journal} {2D Mater.}\ }\textbf {\bibinfo
  {volume} {3}},\ \bibinfo {pages} {015010} (\bibinfo {year}
  {2016})}\BibitemShut {NoStop}%
\bibitem [{\citenamefont {Messina}\ and\ \citenamefont
  {Antezza}(2014)}]{Messina2014}%
  \BibitemOpen
  \bibfield  {author} {\bibinfo {author} {\bibfnamefont {R.}~\bibnamefont
  {Messina}}\ and\ \bibinfo {author} {\bibfnamefont {M.}~\bibnamefont
  {Antezza}},\ }\bibfield  {title} {\bibinfo {title} {Three-body radiative heat
  transfer and casimir-lifshitz force out of thermal equilibrium for arbitrary
  bodies},\ }\href {https://doi.org/10.1103/PhysRevA.89.052104} {\bibfield
  {journal} {\bibinfo  {journal} {Phys. Rev. A}\ }\textbf {\bibinfo {volume}
  {89}},\ \bibinfo {pages} {052104} (\bibinfo {year} {2014})}\BibitemShut
  {NoStop}%
\bibitem [{\citenamefont {Messina}\ and\ \citenamefont
  {Antezza}(2011)}]{Messina2011PRA}%
  \BibitemOpen
  \bibfield  {author} {\bibinfo {author} {\bibfnamefont {R.}~\bibnamefont
  {Messina}}\ and\ \bibinfo {author} {\bibfnamefont {M.}~\bibnamefont
  {Antezza}},\ }\bibfield  {title} {\bibinfo {title} {Scattering-matrix
  approach to casimir-lifshitz force and heat transfer out of thermal
  equilibrium between arbitrary bodies},\ }\href
  {https://doi.org/10.1103/PhysRevA.84.042102} {\bibfield  {journal} {\bibinfo
  {journal} {Phys. Rev. A}\ }\textbf {\bibinfo {volume} {84}},\ \bibinfo
  {pages} {042102} (\bibinfo {year} {2011})}\BibitemShut {NoStop}%
\bibitem [{\citenamefont {Palik}(1998)}]{Palik}%
  \BibitemOpen
  \bibfield  {author} {\bibinfo {author} {\bibfnamefont {E.}~\bibnamefont
  {Palik}},\ }\href@noop {} {\emph {\bibinfo {title} {Handbook of Optical
  Constants of Solids}}}\ (\bibinfo  {publisher} {Academic},\ \bibinfo
  {address} {New York},\ \bibinfo {year} {1998})\BibitemShut {NoStop}%
\bibitem [{\citenamefont {Gu\'erout}\ \emph {et~al.}(2012)\citenamefont
  {Gu\'erout}, \citenamefont {Lussange}, \citenamefont {Rosa}, \citenamefont
  {Hugonin}, \citenamefont {Dalvit}, \citenamefont {Greffet}, \citenamefont
  {Lambrecht},\ and\ \citenamefont {Reynaud}}]{Greffet2012prb_trans_def}%
  \BibitemOpen
  \bibfield  {author} {\bibinfo {author} {\bibfnamefont {R.}~\bibnamefont
  {Gu\'erout}}, \bibinfo {author} {\bibfnamefont {J.}~\bibnamefont {Lussange}},
  \bibinfo {author} {\bibfnamefont {F.~S.~S.}\ \bibnamefont {Rosa}}, \bibinfo
  {author} {\bibfnamefont {J.-P.}\ \bibnamefont {Hugonin}}, \bibinfo {author}
  {\bibfnamefont {D.~A.~R.}\ \bibnamefont {Dalvit}}, \bibinfo {author}
  {\bibfnamefont {J.-J.}\ \bibnamefont {Greffet}}, \bibinfo {author}
  {\bibfnamefont {A.}~\bibnamefont {Lambrecht}},\ and\ \bibinfo {author}
  {\bibfnamefont {S.}~\bibnamefont {Reynaud}},\ }\bibfield  {title} {\bibinfo
  {title} {Enhanced radiative heat transfer between nanostructured gold
  plates},\ }\href {https://doi.org/10.1103/PhysRevB.85.180301} {\bibfield
  {journal} {\bibinfo  {journal} {Phys. Rev. B}\ }\textbf {\bibinfo {volume}
  {85}},\ \bibinfo {pages} {180301(R)} (\bibinfo {year} {2012})}\BibitemShut
  {NoStop}%
\bibitem [{\citenamefont {Liu}\ and\ \citenamefont
  {Zhang}(2014)}]{Liu2014apl_trans_def}%
  \BibitemOpen
  \bibfield  {author} {\bibinfo {author} {\bibfnamefont {X.~L.}\ \bibnamefont
  {Liu}}\ and\ \bibinfo {author} {\bibfnamefont {Z.~M.}\ \bibnamefont
  {Zhang}},\ }\bibfield  {title} {\bibinfo {title} {{Graphene-assisted
  near-field radiative heat transfer between corrugated polar materials}},\
  }\href {https://doi.org/10.1063/1.4885396} {\bibfield  {journal} {\bibinfo
  {journal} {Appl. Phys. Lett.}\ }\textbf {\bibinfo {volume} {104}},\ \bibinfo
  {pages} {251911} (\bibinfo {year} {2014})}\BibitemShut {NoStop}%
\bibitem [{\citenamefont {Zhou}\ \emph
  {et~al.}(2022{\natexlab{b}})\citenamefont {Zhou}, \citenamefont {Zhang},
  \citenamefont {Torbatian}, \citenamefont {Novko}, \citenamefont {Antezza},\
  and\ \citenamefont {Yi}}]{Zhou2022prm}%
  \BibitemOpen
  \bibfield  {author} {\bibinfo {author} {\bibfnamefont {C.-L.}\ \bibnamefont
  {Zhou}}, \bibinfo {author} {\bibfnamefont {Y.}~\bibnamefont {Zhang}},
  \bibinfo {author} {\bibfnamefont {Z.}~\bibnamefont {Torbatian}}, \bibinfo
  {author} {\bibfnamefont {D.}~\bibnamefont {Novko}}, \bibinfo {author}
  {\bibfnamefont {M.}~\bibnamefont {Antezza}},\ and\ \bibinfo {author}
  {\bibfnamefont {H.-L.}\ \bibnamefont {Yi}},\ }\bibfield  {title} {\bibinfo
  {title} {Photon tunneling reconstitution in black phosphorus/$\mathrm{hBN}$
  heterostructure},\ }\href {https://doi.org/10.1103/PhysRevMaterials.6.075201}
  {\bibfield  {journal} {\bibinfo  {journal} {Phys. Rev. Mater.}\ }\textbf
  {\bibinfo {volume} {6}},\ \bibinfo {pages} {075201} (\bibinfo {year}
  {2022}{\natexlab{b}})}\BibitemShut {NoStop}%
\bibitem [{\citenamefont {Noto}\ \emph {et~al.}(2014)\citenamefont {Noto},
  \citenamefont {Messina}, \citenamefont {Guizal},\ and\ \citenamefont
  {Antezza}}]{Casimir_Noto}%
  \BibitemOpen
  \bibfield  {author} {\bibinfo {author} {\bibfnamefont {A.}~\bibnamefont
  {Noto}}, \bibinfo {author} {\bibfnamefont {R.}~\bibnamefont {Messina}},
  \bibinfo {author} {\bibfnamefont {B.}~\bibnamefont {Guizal}},\ and\ \bibinfo
  {author} {\bibfnamefont {M.}~\bibnamefont {Antezza}},\ }\bibfield  {title}
  {\bibinfo {title} {Casimir-lifshitz force out of thermal equilibrium between
  dielectric gratings},\ }\href {https://doi.org/10.1103/PhysRevA.90.022120}
  {\bibfield  {journal} {\bibinfo  {journal} {Phys. Rev. A}\ }\textbf {\bibinfo
  {volume} {90}},\ \bibinfo {pages} {022120} (\bibinfo {year}
  {2014})}\BibitemShut {NoStop}%
\bibitem [{\citenamefont {Wang}\ \emph {et~al.}(2021)\citenamefont {Wang},
  \citenamefont {Tang}, \citenamefont {Ng}, \citenamefont {Messina},
  \citenamefont {Guizal}, \citenamefont {Crosse}, \citenamefont {Antezza},
  \citenamefont {Chan},\ and\ \citenamefont {Chan}}]{Hobun}%
  \BibitemOpen
  \bibfield  {author} {\bibinfo {author} {\bibfnamefont {M.}~\bibnamefont
  {Wang}}, \bibinfo {author} {\bibfnamefont {L.}~\bibnamefont {Tang}}, \bibinfo
  {author} {\bibfnamefont {C.~Y.}\ \bibnamefont {Ng}}, \bibinfo {author}
  {\bibfnamefont {R.}~\bibnamefont {Messina}}, \bibinfo {author} {\bibfnamefont
  {B.}~\bibnamefont {Guizal}}, \bibinfo {author} {\bibfnamefont {J.~A.}\
  \bibnamefont {Crosse}}, \bibinfo {author} {\bibfnamefont {M.}~\bibnamefont
  {Antezza}}, \bibinfo {author} {\bibfnamefont {C.~T.}\ \bibnamefont {Chan}},\
  and\ \bibinfo {author} {\bibfnamefont {H.~B.}\ \bibnamefont {Chan}},\
  }\bibfield  {title} {\bibinfo {title} {Strong geometry dependence of the
  casimir force between interpenetrated rectangular gratings},\ }\href
  {https://doi.org/10.1038/s41467-021-20891-4} {\bibfield  {journal} {\bibinfo
  {journal} {Nat. Commun.}\ }\textbf {\bibinfo {volume} {12}},\ \bibinfo
  {pages} {600} (\bibinfo {year} {2021})}\BibitemShut {NoStop}%
\bibitem [{\citenamefont {Hwang}(2020{\natexlab{b}})}]{Hwang2020}%
  \BibitemOpen
  \bibfield  {author} {\bibinfo {author} {\bibfnamefont {R.-B.}\ \bibnamefont
  {Hwang}},\ }\bibfield  {title} {\bibinfo {title} {Highly improved convergence
  approach incorporating edge conditions for scattering analysis of graphene
  gratings},\ }\href {https://doi.org/10.1038/s41598-020-69827-w} {\bibfield
  {journal} {\bibinfo  {journal} {Sci. Rep.}\ }\textbf {\bibinfo {volume}
  {10}},\ \bibinfo {pages} {12855} (\bibinfo {year}
  {2020}{\natexlab{b}})}\BibitemShut {NoStop}%
\end{thebibliography}%

\end{document}